\documentclass[twocolumn,dvipsnames]{aastex7}

\usepackage{graphicx}
\usepackage{amssymb,amsmath}
\usepackage{bbm}
\usepackage{mathrsfs} 
\usepackage{latexsym}
\usepackage{mathtools} 
\usepackage{color}
\usepackage{comment}
\usepackage[normalem]{ulem} 
\usepackage{dcolumn}
\usepackage[acronym]{glossaries}
\usepackage{algpseudocode}
\usepackage{algorithm}
\usepackage{booktabs}
\usepackage{xspace}

\let\listofchanges\undefined

\usepackage[deletedmarkup=xout,commentmarkup=uwave,commandnameprefix=ifneeded]{changes}

\renewcommand{\added}{\chadded}

\definechangesauthor[name=Max, color=teal]{MI}
\definechangesauthor[name=Aaron, color=purple]{AZ}
\definechangesauthor[name=Asad, color=brown]{AH}

\allowdisplaybreaks

\graphicspath{{.}}

\newcommand{\Austin}{\affiliation{Weinberg Institute, University of Texas at Austin, Austin, TX 78712, USA}}
\newcommand{\CCA}{\affiliation{Center for Computational Astrophysics, Flatiron Institute, NY}}

\newcommand{\btheta}{\boldsymbol{\theta}}
\newcommand{\bmu}{\boldsymbol{\mu}}
\newcommand{\bSigma}{\boldsymbol{\Sigma}}
\newcommand{\bLambda}{\boldsymbol{\Lambda}}
\newcommand{\ba}{{\bf a}}
\newcommand{\bb}{{\bf b}}

\newcommand{\bTheta}{\boldsymbol{\Theta}}

\newcommand{\R}[1]{\textcolor{red}{#1}}

\newcommand{\grad}{\nabla}
\newcommand{\one}[1]{\mathds{1}\left[#1\right]}

\newcommand{\sigmanoise}{\sigma_{\textrm{noise}}}
\newcommand{\pdet}{P_{\text{det}}}

\renewcommand{\d}{\mathrm{d}}

\newacronym{GW}{GW}{gravitational wave}
\newacronym{GR}{GR}{general relativity}
\newacronym{CBC}{CBC}{compact binary coalescence}
\newacronym{BH}{BH}{black hole}
\newacronym{BBH}{BBH}{binary black hole}
\newacronym{LVK}{LVK}{LIGO-Virgo-KAGRA}
\newacronym{PE}{PE}{parameter estimation}
\newacronym{FAR}{FAR}{false-alarm rate}
\newacronym{GWOSC}{GWOSC}{the Gravitational Wave Open Science Center}
\newacronym{SSB}{SSB}{solar system barycenter}
\newacronym{SPA}{SPA}{stationary phase approximation}
\newacronym{PN}{PN}{post-Newtonian}
\newacronym{BNS}{BNS}{binary neutron star}
\newacronym{IMR}{IMR}{inspiral-merger-ringdown}
\newacronym{NSF}{NSF}{National Science Foundation}
\newacronym{MC}{MC}{Monte-Carlo}
\newacronym{GMM}{GMM}{Gaussian mixture model}
\newacronym{MCMC}{MCMC}{Markov chain Monte Carlo}
\newacronym{BIC}{BIC}{Bayesian information criterion}
\newacronym{BF}{BF}{Bayes factor}
\newacronym{SDDR}{SDDR}{Savage Dickey density ratio}
\newacronym{TGMM}{TGMM}{truncated Gaussian mixture model}
\newacronym{PPD}{PPD}{posterior predictive distribution}
\newacronym{HPDI}{HPDI}{highest posterior density interval}
\newacronym{AGN}{AGN}{active galactic nucleus}
\newacronym{KDE}{KDE}{kernel density estimate}
\newacronym{IID}{IID}{identically and independently distributed}
\newacronym{PDF}{PDF}{probability density function}

\glsdisablehyper

\begin{document}
	
		\title{Living on the edge: Testing for compact population features at the edges of parameter space}
	
	\author[orcid=0000-0003-3491-5439]{Asad Hussain}
	\CCA
	\Austin
	\email[show]{asadh@utexas.edu}

	\author[orcid=0000-0001-8830-8672]{Maximiliano Isi}
	\affiliation{Department of Astronomy, Columbia University, New York, NY 10027, USA}
	\CCA
	\email{}

	\author[orcid=0000-0002-7453-6372]{Aaron Zimmerman}
	\Austin
	\email{}

	\date{\today}

	\newcommand{\gravpop}{\textsc{gravpop}\xspace}
	\newcommand{\truncatedgaussianmixtures}{\textsc{TruncatedGaussianMixtures}\xspace}
	
	\begin{abstract}
		Many astrophysical population studies involve parameters that exist on a bounded domain, such as the dimensionless spins of black holes or the eccentricities of planetary orbits, both of which are confined to $[0, 1]$.
		In such scenarios, we often wish to test for distributions clustered near a boundary, e.g., vanishing spin or orbital eccentricity.
		Conventional approaches---whether based on Monte Carlo, kernel density estimators, or machine-learning techniques---often suffer biases at the boundaries. 
		These biases stem from sparse sampling near the edge, kernel-related smoothing, or artifacts introduced by domain transformations. 
		We introduce a truncated Gaussian mixture model framework that substantially mitigates these issues, enabling accurate inference of narrow, edge-dominated population features.
		While our method has broad applications to many astronomical domains, we consider gravitational wave catalogs as a concrete example to demonstrate its power.
		In particular, we maintain agreement with published constraints on the fraction of zero-spin binary black hole systems in the GWTC-3 catalog—--results originally derived at much higher computational cost through dedicated reanalysis of individual events in the catalog.
		Our method can achieve similarly reliable results with a much lower computational cost.
		The method is publicly available in the open-source packages \gravpop and \truncatedgaussianmixtures.
	\end{abstract}
	
	\section{Introduction} \label{sec:intro}

	Modern astrophysical surveys produce ever-growing catalogs of objects including binary systems, exoplanets, and compact objects, allowing us to probe their underlying formation and evolutionary processes. 
	These surveys often report measurements of the properties of individual objects and their corresponding uncertainties, from which global properties of the distribution of astrophysical sources can be inferred.
	This is best achieved via hierarchical Bayesian inference---as is done for, e.g., \gls{GW} sources \citep{theligoscientificcollaboration2025gwtc40populationpropertiesmerging, theligoscientificcollaboration2025gwtc40updatinggravitationalwavetransient, theligoscientificcollaboration2025gwtc40methodsidentifyingcharacterizing, KAGRA:2023pio, LIGOScientific:2014pky}, exoplanet populations \citep{hogg_inferring_2010, rogers_most_2015, wolfgang_probabilistic_2016, lustig-yaeger_hierarchical_2022, keating_atmospheric_2021, cloutier2024exoplanetl}, or stellar populations \citep{Leistedt_2023}.

	Although efficient techniques have been developed to carry out hierarchical inference in different contexts, they often have difficulties with bounded domains.
	This is the case for the magnitudes of \gls{BH} spins $\chi$ measured using \glspl{GW} or the eccentricities $\epsilon$ of exoplanets inferred from transit data, both of which are confined to the unit interval.
	With existing techniques, it is difficult to answer questions like \textit{``Do all black holes have negligible spin?''} or \textit{``What is the distribution of non-eccentric exoplanets?''}, which require testing population distributions with sharp features at the boundary of the domain (i.e., $\chi = 0$ or $\epsilon = 0$).

	In this paper, we propose a new method for hierarchical inference in bounded domains, improving upon traditional techniques both in efficiency and accuracy.
	We achieve this through a framework based on \glspl{TGMM} that allows us to accurately estimate the hierarchical likelihood even under populations with sharp features at the boundary of the domain.
	We show that this method does not introduce additional bias, and present efficient procedures for implementing it.
	Although our method is general, we focus on modeling the population of \gls{BH} spin magnitudes from \gls{LVK} \gls{GW} data as a concrete application.
	A novel application of this method to understand the structure in the distribution of \gls{BBH} spins using gravitational wave detections from the GWTC-3 catalog~\citep{KAGRA:2021vkt} can be found in our companion paper \citep{hussainHintsSpinmagnitudeCorrelations2024a}.

	This paper is organized as follows.
	In Sec.~\ref{sec:background}, we introduce the hierarchical inference problem in the context of \glspl{GW}.
	In Sec.~\ref{sec:GeneralProcedure}, we present our method in full generality and compare it to current techniques.
	In Sec.~\ref{sec:GWApplication}, we apply our method to the population analysis of \gls{BBH} mergers detected by the \gls{LVK} in the GWTC-3 catalog~\citep{KAGRA:2021vkt}.
	In Sec.~\ref{sec:ZeroSpin}, we revisit the constraints from \cite{Tong2022}\footnote{We note that \cite{Adamcewicz2025-hm} points out a possible error in the analysis of \cite{Tong2022}.} on the fraction of zero-spin \glspl{BBH} in the GWTC-3 catalog.
	In Sec.~\ref{sec:Conclusions}, we conclude and discuss further applications of our methods.
	The appendices contain additional details of our derivations and methods.
	
	\section{Background: GW inference}
	\label{sec:background}

	The problem of estimating the population of astrophysical parameters near a boundary has surfaced with high prominence in the context of spin measurements in \gls{GW} astronomy.
	Here we summarize the nature of the technical problem in this context.
	
	Population inference of \glspl{BBH} detected by the \gls{LVK} begins with the posterior probability distributions inferred from the data for each detection.
	These posteriors are typically represented by discrete sets of samples drawn through stochastic algorithms such as \gls{MCMC} \citep{metropolisEquationStateCalculations1953, foreman-mackeyEmceeMCMCHammer2013} or nested sampling \citep{skillingNestedSamplingGeneral2006}. 
	To evaluate the degree to which a proposed astrophysical population model explains the data, one must compute the Bayesian evidence for each event under a prior corresponding to that model \citep[see, e.g.,][]{Thrane:2018qnx}.

	Rather than recomputing the single-event posteriors for each prior, the evidence is generally estimated by reweighting preexisting posterior samples: each sample is assigned a new weight proportional to the ratio between the proposed prior and the fiducial prior used in the original event-level analysis \citep{Thrane:2018qnx}.
	Additionally, it is standard in \gls{GW} astronomy to account for the Malmquist selection bias using injection campaigns \citep[see, e.g.,][for more details]{essick_semianalytic_2023}.
	This requires a similar reweighting step \citep{Farr:2019rap}, this time to replace the fiducial distribution used to draw injections with the desired astrophysical one.
	Both these calculations suffer from similar difficulties in dealing with sharp population features.

	The challenges in assessing population features near domain boundaries begin with problems intrinsic to sharp, localized features in general.
	It is well documented that the reweighting approach described above fails for sharp features \citep{Talbot:2020oeu, Talbot2023}. 
	The reason for this is straightforward: assuming a broad posterior, it is unlikely that sufficient samples are found in the compact region where the feature lies to inform the estimate accurately \citep{robert_monte_2004}.
	
	There have been many proposals to handle sharp population features within the context of \gls{GW} data analysis. \citep[see, e.g.,][to name a few]{wysocki2020inferringneutronstarequation, Lackey_2015, Carney_2018, delfavero2022compressedparametricnonparametricapproximations, delfavero2022normalapproximatelikelihoodsgravitational, Callister:2022qwb, Talbot:2020oeu, Callister:2024qyq, mancarella2025samplinghierarchicalpopulationposterior}
	\cite{Callister:2022qwb} use one-dimensional Gaussian \glspl{KDE} to represent the posterior in the effective spin parameter $\chi_{\textrm{eff}}$, and use that to interpolate between samples and probe sharp features in this domain.
	This is facilitated by properties of the Gaussian distribution, which can be leveraged to perform some of the computation analytically and behaves well even for narrow features.
	Similarly \cite{Talbot:2020oeu} and \cite{Callister:2024qyq} respectively use \glspl{GMM} and neural networks to estimate densities from injection campaigns. 
	Unlike the \glspl{KDE} with analytic integration, these approaches rely on generating samples from the new population prior, and then using the density estimates to reweight these samples. 
	For sharp population features this can be an efficient estimator of the evidence under the new prior.
	In general, the goal of such density estimation techniques is to get an accurate representation of the distribution that makes it computationally efficient to estimate the evidence integral.
	
	However, such density estimation techniques typically fail to capture sharp population features near the edges of a domain.
	In fact, most approaches either ignore the existence of a boundary or rely on transforming bounded parameters to unbounded domains.
	In both cases, the density estimates introduce inherent biases at the boundary, which can vary in severity. 
	A Gaussian \gls{KDE}, for example, would have a bias at the boundary which is $\mathcal{O}(1)$ in the bandwidth $h$, while having a bias of order $\mathcal{O}(h^2)$ in the bulk \citep{silvermanDensityEstimationStatistics2018}.
	
	An ideal density estimation method would fulfill two requirements. Firstly, it should not give biased estimates of the \gls{PDF} at or near the edges. Secondly, the representation of the density should facilitate the evidence computation, making it fast, tractable, and accurate. 
	In this paper, we find that utilizing \glspl{TGMM} fulfills the above criteria.
	
	\section{General Procedure}
	\label{sec:GeneralProcedure} 
	The main target of our framework is the accurate and fast estimation of integrals of the form
	\begin{equation}
		\label{eq:the-generic-integral}
		I(\bLambda) = \int p(\btheta | \bLambda) \frac{p(\btheta)}{W(\btheta)} \d^n\theta = \left\langle   \frac{p(\btheta|\Lambda)}{W(\btheta)}  \right\rangle_{\btheta \sim p(\btheta)}\, ,
	\end{equation} 
	given some set of parameter samples $\btheta_i \sim p(\btheta)$ and some corresponding set of weights given by $W_i = W(\btheta_i)$.
	The target probability distribution is defined by population-level parameters $\bLambda$. 
	The notation $\langle f(\btheta) \rangle_{\btheta \sim p(\btheta)}$ represents the expectation of function $f(\btheta)$ under the distribution $p(\btheta)$.

	These integrals arise when reweighting posterior samples or injection draws for the selection function.
	Below, we show how the integrals appear, describe how to compute them with our method, and compare it to standard procedures used in \gls{GW} astronomy.

	\subsection{Formal problem statement and notation}
	
	We analyze a catalog of $N$ observed objects (detection ``events'' in the \gls{GW} context), indexed by $e=1,\dots,N$.  
	Each object yields a data set $d_e$ (e.g., a \gls{GW} strain time series), collectively creating a data catalog $\mathcal D = \{d_e ~\forall~ e \in 1\dots N\}$. 
	Initial parameter estimation applied to each object produces samples drawn from the posterior 
	\begin{equation}
		p(\btheta \mid d_e)  = \frac{\mathcal{L}(d_e \mid \btheta)\, \pi(\btheta \mid \emptyset)}{\mathcal{Z}(d_e)}\\,
	\end{equation}
	where $\btheta$ parameterizes the properties of a single event. This posterior is produced by assuming a fiducial prior $\pi(\btheta\mid\emptyset)$ called the \emph{sampling prior}, and a model for the data generation process $\mathcal{L}(d_e \mid \btheta)$ hereby called the \emph{likelihood}; additionally, $\mathcal{Z}(d_e)$ is the \emph{evidence} under the sampling prior, which acts as a normalization factor.
	
	For example, in \gls{GW} \gls{BBH} population analyses, one commonly considers the mass of the heavier black hole $m_1$, the ratio of the masses in the binary $q$, the cosine of the individual spin tilts $\theta_i$, the individual spin magnitudes $\chi_i$, and the redshift $z$ at which the event took place,\footnote{While there are other parameters that are usually inferred, we usually analyze only a subset when modeling the population.}
	\begin{equation} \label{eq:BBH-parameters}
		\btheta_{\text{BBH}} = [m_1, q, \cos\theta_1, \cos\theta_2, \chi_1, \chi_2, z]\,.
	\end{equation}
	In this case, $\cos\theta_i$ is bounded within $[-1,1]$ and $\chi_i$ is bounded within $[0,1]$.
	The cases where $\chi_i \to 0$ or $\cos\theta_i \to 1$ are physically interesting and it is desirable to explore population features which have compact support at these boundaries. 
	
	After obtaining posterior samples for each object, we create a model for the population distribution of astrophysical objects $p(\btheta \mid \bLambda)$, parameterized by some hyperparameters $\bLambda$. 
	The goal is to infer the posterior over the parameters of these distributions, given our detected catalog $\mathcal D$, written as $p(\bLambda \mid \mathcal D)$.
	To do this, we set some prior over our hyperparameters, $\pi(\bLambda)$.
	Under suitable assumptions \citep{Essick:2023upv}, the population posterior can then be written as,
	\begin{equation}
		p(\bLambda \mid \mathcal D) \propto \pi(\bLambda)\, \xi(\bLambda)^{-N}\prod_{e}^N \mathcal{L}(d_e \mid \bLambda)\,,
		\label{eq:populationposterior}
	\end{equation}
	with $\mathcal{L}(d_e \mid \bLambda)$ and $\xi(\bLambda)$ as defined below. 
	
	The $\mathcal{L}(d_e \mid \bLambda)$ factors represent the evidence for each event given by data $d_e$, under a prior defined by the chosen astrophysical population distribution $p(\btheta \mid \bLambda)$ instead of the sampling prior $\pi(\btheta \mid \emptyset)$.
	This is typically estimated from the initial set of posterior samples as
	\begin{equation}
		\mathcal{L}(d_e \mid \bLambda) =  \int p(\btheta \mid \bLambda) \frac{p(\btheta\mid d_e)}{\pi(\btheta \mid \emptyset)} \d\btheta \,.
		\label{eq:eventlevelposterior}
	\end{equation}
	We typically do not have direct access to $p(\btheta\mid d_e)$ but instead have a discrete set of independent samples $\{\btheta_{e,j}\}$ drawn from it, where $j$ ranges from 1 to $N_e$, the number of posterior samples for event $e$.
	The integral is then approximated via a Monte Carlo average, as
	\begin{equation}
		\mathcal{L}(d_e \mid \bLambda) \approx \frac{1}{N_e} \sum_{j=1}^{N_e} \frac{p(\btheta_{e,j} \mid \bLambda)}{\pi(\btheta_{e,j} \mid \emptyset)} \,.
		\label{eq:eventlevelposterior_mc}
	\end{equation}
	
	The $\xi(\bLambda)$ factor is the population-averaged detection efficiency, which accounts for selection effects \citep{Mandel:2018mve}.
	Since the detection process is imperfect, the detectability of different objects varies over parameter space.
	The probability of detecting an object with parameters $\btheta$ is denoted $P_{\text{det}}(\btheta)\in(0,1)$. 
	The fraction of events in the population distribution $p(\btheta\mid\bLambda)$ that are expected to be detected is given by
	\begin{equation}
		\xi(\boldsymbol{\Lambda}) = \int p(\boldsymbol{\theta} \mid \bLambda)\,  P_{\rm det}(\boldsymbol{\theta})\, \d\boldsymbol{\theta}\,.
		\label{eq:detectionefficiency0}
	\end{equation}
	We typically do not have direct access to $P_{\rm det}(\btheta)$ but must estimate it through injection campaigns.
	This entails drawing samples from a fiducial population with broad support, $p(\btheta \mid \bLambda_\emptyset)$, and generating synthetic data for each sample.
	These synthetic data are passed through the same detection pipelines used to construct the catalog $\{d_e\}$, with each injection labeled as detected or not detected.
	The detection efficiency can then be recast as a Monte Carlo average.
	Given $N_{\rm inj}$ injected signals drawn from a reference distribution $p_{\rm inj}(\btheta)$, the detection efficiency is approximated by
	\begin{equation}
		\xi(\boldsymbol{\Lambda}) \approx \frac{1}{N_{\rm inj}} \sum_{i=1}^{N_{\rm found}} \frac{p(\boldsymbol{\theta}_i \mid \boldsymbol{\Lambda})}{p_{\rm inj}(\boldsymbol{\theta}_i)}\,,
		\label{eq:detectionefficiency}
	\end{equation}
	where the sum is over $N_{\rm found}$ injections that pass the detection criteria.
	
	The fundamental assumption underlying Eqs.~\eqref{eq:eventlevelposterior_mc} and \eqref{eq:detectionefficiency} is that the posterior samples or recovered injections cover the feature of interest sufficiently densely.
	This assumption may be violated for narrow features in the population distribution, such as sharp peaks or features near parameter boundaries.
	
	In summary, the main ingredients of a typical hierarchical inference procedure, Eqs.~\eqref{eq:eventlevelposterior_mc} and \eqref{eq:detectionefficiency}, take the form of Eq.~\eqref{eq:the-generic-integral}.
	This is the problem we aim to tackle for narrow population features close to the edge.

	\subsection{Common techniques for approximating inference integrals}
	\label{sec:techniques-for-computing}

	Here we review three common strategies for estimating Eq.~\eqref{eq:the-generic-integral} starting from samples of $p(\btheta)$: a \gls{MC} estimator, a regular \gls{KDE}, and a \gls{KDE} with reflective boundary conditions.
	This is not an exhaustive list, even within the context of gravitational waves, but is meant to be illustrative.

\textbf{Monte-Carlo:} The most common approach for approximating Eq.~\eqref{eq:the-generic-integral} is to use a \gls{MC} estimator given by
\begin{equation}
	\hat{I}^{MC}(\bLambda) \approx \frac{1}{N}\sum_i^N \frac{p(\btheta_i \mid \bLambda)}{W(\btheta_i)}\, ,
\end{equation}
such as Eqs.~\eqref{eq:eventlevelposterior_mc} and \eqref{eq:detectionefficiency}.
This is the method of choice in \gls{LVK} population analyses \citep{LVKPop}.
The \gls{MC} estimator can be shown to be asymptotically unbiased: given enough samples it will converge to the correct value of the integral. However, the estimator can have a very large variance for sharp features \citep{robert_monte_2004}.

\textbf{Kernel Density Estimation:} Density estimation techniques introduce some level of bias in order to tame the excessive variance of the estimator in the regime of sharp population features.
At minimum, this typically includes regularizing the integral by imposing smoothness on the \gls{PDF}.
This class of estimators begins by constructing a smooth approximation to $p(\btheta)$ using a kernel function. 

A kernel density estimate (KDE) is built by centering a kernel function, $K_h(\btheta, \btheta_i)$, at each sample $\btheta_i$, and summing over all samples:
\begin{equation}
	p(\btheta) \approx \frac{1}{N} \sum_{i=1}^N K_h(\btheta, \btheta_i)\,.
\end{equation}
Here, $K_h$ is typically a normalized, symmetric function (such as a Gaussian) with a parameter $h$ called the bandwidth, which controls the width of the kernel and thus the amount of smoothing. The choice of bandwidth is crucial: a small $h$ leads to less smoothing and can capture sharp features but may introduce noise, while a large $h$ smooths out noise but can wash out narrow features. Bandwidth can be set using rules of thumb or cross-validation techniques \citep{crossvalidation-KDE, silvermanDensityEstimationStatistics2018}.

For example, consider targeting narrow features in the redshift $z$ distribution of \gls{BBH}s.
Let us break the parameter space as $\btheta = [\btheta', z]$ where $\btheta'$ are all the remaining parameters describing the event. Similar to \cite{Callister:2022qwb}, we can represent $p(\btheta)$ through a \gls{KDE}---applying kernel smoothing only in the subspace of interest (the redshift in this example), and using point estimates (delta functions) in the remainder of parameter space.
The density estimate would be,
\begin{equation}
	\label{eq:kde-estimate-redshift}
	p(\btheta) \approx \frac{1}{N}\sum_{i=1}^N \delta(\btheta' - \btheta'_i)\, \phi(z \mid z_i, h)\,,
\end{equation}
where $\phi(z \mid z_i, h)$ is a Gaussian kernel centered at $z_i$ with bandwidth $h$.

Similarly, we may apply a population model such that
\begin{equation} \label{eq:population-model-redshift}
p(\btheta \mid \bLambda) = p(\btheta' \mid \bLambda')\, \phi(z \mid \mu_z, \sigma_z)\,,
\end{equation}
setting the population model in the redshift sector to be a Gaussian with some mean $\mu_z$ and width $\sigma_z$, while applying some arbitrary population model to the other parameters $\btheta'$.
One could then plug Eqs.~\eqref{eq:kde-estimate-redshift} and \eqref{eq:population-model-redshift} into Eq.~\eqref{eq:the-generic-integral} and obtain an estimate of the integral as,
\begin{align}
	\label{eq:KDE-integral}
	\hat{I}^{KDE}(\bLambda) &= \frac{1}{N}\sum_{i=1}^N \left[ \frac{p(\btheta'_i \mid \bLambda)}{W(\btheta_i)} \right. \nonumber \\
	&\left. \times \int \phi(z \mid z_i, h) \phi(z \mid \mu_z, \sigma_z){\rm d}z \right] \nonumber \\
	&\approx \frac{1}{N}\sum_{i=1}^N \frac{p(\btheta'_i \mid \bLambda)}{W(\btheta_i)} F(z_i, h, \mu_z, \sigma_z)\, ,
\end{align}
where $F$ is the result of the convolution of the two Gaussians, which has an analytic form that is quick to evaluate.
Since we compute $F$ analytically, we can get good estimates of the integral even as the shape of the population model becomes sharp (i.e., $\sigma_z \to 0$).
If the kernel and population model were not both Gaussian, $F$ may not have an analytic form, and the benefit of this approach would be reduced.

The bias properties of the integral estimator $\hat{I}^{KDE}(\bLambda)$ are only as good as that of the \gls{KDE} estimate itself.
The bias of the density estimate using a Gaussian \gls{KDE} built from $n$ samples in $d$ dimensions and bandwidth $h$ is known to be low in the bulk of parameter space, $\mathcal{O}(h^2)$ \citep{silvermanDensityEstimationStatistics2018}; this is especially true for low dimensions because $h \sim n^{-1/d}$
Consequently, the \gls{KDE} approach above is a good estimator of the targeted integral when we stay in the bulk.
However, at the boundary, the bias deviates from the $\mathcal{O}(h^2)$ in the bulk and becomes $\mathcal{O}(1)$ in the case of the standard Gaussian \gls{KDE} \citep{silvermanDensityEstimationStatistics2018}. This means that the value at the boundary is always incorrect regardless of the number of samples used. 

\textbf{Reflective \gls{KDE}s: }There are ways to improve the bias of the \gls{KDE} estimators at the boundary. For instance, a common strategy is to reflect the points about the boundary, and have them inform the estimate of the density inside the bounded region \citep{schusterIncorporatingSupportConstraints1985}.
In the redshift example above, assuming a bounded $z$ parameter $z \in [a,b]$, the reflective \gls{KDE} would take the form
\begin{align}
		\label{eq:reflective-kde-estimate-redshift}
	p(\btheta) &\approx \frac{1}{N}\sum_{i}^N \delta(\btheta' - \btheta'_i)\left[ \phi(z \mid z_i\,, h)  \right. \nonumber \\ 
	  &\left. +\, \phi(z \mid 2a - z_i\,, h) + \phi(z \mid 2b - z_i\,, h)\right]\,.
\end{align}
This improves the bias of the \gls{KDE} to $\mathcal{O}(h)$ near the boundary while retaining the smaller $\mathcal{O}(h^2)$ bias in the bulk.
The reflective \gls{KDE} produces an expression for the estimator of the integral similar to Eq.~\eqref{eq:KDE-integral},
\begin{align}
	\label{eq:reflective-KDE-integral}
	\hat{I}^{RKDE}(\bLambda) 
	&\approx \frac{1}{N}\sum_{i}^N \frac{p(\btheta'_i \mid \bLambda)}{W(\btheta_i)} \big[ F(z_i, h, \bLambda_z) \nonumber \\
    &+  F(a - z_i, h, \bLambda_z) + F(2b - z_i, h, \bLambda_z)\big]\,.
\end{align}
However, if the derivative at the boundary is non-zero, as is generically the case, this may lead to critical biases \citep{marronTransformationsReduceBoundary1994}.

\subsection{Numerical Comparison of Methods}
\label{sec:numerical-comparison-of-methods}


	To explore the different estimators of Eq.~\eqref{eq:the-generic-integral} used in \gls{GW} population analysis we analyze their bias and variance properties for a simple 1D toy population analysis. 
%
	%
	This toy model demonstrates how the above \gls{MC} and \gls{KDE} methods break down when the population prior is sharply peaked at a boundary and what a robust estimator must accomplish instead.
	
	We model the population of a single parameter $\chi$ that is bounded to the unit interval, $\chi \in [0,1]$, as coming from a truncated normal distribution,
	\begin{equation}
		\label{spins_are_truncated_normal}
		\chi \sim \mathcal{N}_{[0,1]}(\mu, \sigma)\, ,
	\end{equation}
	with mean $\mu$ and standard deviation $\sigma$.
	We also use this distribution to draw a true value for the parameter $\chi^{{\rm true}}_e$ for each event, i.e., $\chi^{{\rm true}}_e \sim \mathcal{N}_{[0,1]}(\mu, \sigma)$.

	For each event, we assume we have noisy data from which we derive a likelihood that can be approximated as a Gaussian with a width $\sigma_{\text{noise}}$ set by the noise level.
	For the $e$th event, this results in a likelihood represented by a mean $\chi^{{\rm obs}}_e$ and width $\sigma_{\text{noise}}$, such that
	\begin{align}
		\chi^{{\rm obs}}_e &\sim \mathcal{N}(\chi^{{\rm true}}_e, \sigmanoise)\,.
	\end{align}
	The dataset of detections constitute a catalog $\mathcal D = \{\chi^{{\rm obs}}_e\ \forall\ e \in 1,\dots, N\}$ and we assume no selection bias.
	
	When assuming a uniform sampling prior $\pi(\chi) = 1$, the posterior distribution for the $e$th event then takes the form
		\begin{equation}
		p(\chi \mid \chi^{{\rm obs}}_e) = \phi_{[0,1]}(\chi \mid \chi^{{\rm obs}}_e, \sigmanoise)\,,
	\end{equation}
	where $\phi_{[a,b]}(x \mid \mu, \sigma)$ is the probability density function of the truncated normal distribution $\mathcal N_{[a,b]}(\mu,\sigma)$.
	
	We would like to get a posterior distribution over the hyperparameters of the model, $\Lambda = \{\mu, \sigma\}$.
	This can be written as
	\begin{equation}
		p(\mu, \sigma | \{\chi^{{\rm obs}}_e\}) = \pi(\mu, \sigma) \prod_{e} \mathcal{L}(\chi^{{\rm obs}}_e | \mu, \sigma) \, ,
	\end{equation}
	where $ \pi(\mu, \sigma)$ is our prior over the hyper parameters $\mu$ and $\sigma$, and the likelihood factors are given by
	\begin{equation}
		\label{eq:ToyExampleIntegral}
		\mathcal{L}(\chi^{{\rm obs}}_e | \mu, \sigma) = \int \phi_{[0,1]}(\chi | \mu, \sigma)\, \phi_{[0,1]}(\chi | \chi^{{\rm obs}}_e, \sigmanoise) \, \d\chi \, .
	\end{equation}
	This corresponds to Eq.~\eqref{eq:the-generic-integral} with the replacements,
	\begin{subequations}
	\begin{align}
		\btheta &\to \chi \\
		\bLambda &\to \{\mu,\sigma\} \\
		p(\btheta) &\to \phi_{[0,1]}(\chi | \chi^{{\rm obs}}_e, \sigmanoise) \\
		W(\btheta) &\to \pi(\chi) = 1 \\
		p(\btheta \mid \bLambda) &\to  \phi_{[0,1]}(\chi | \mu, \sigma)\\
		I(\bLambda) &\to \mathcal{L}(\chi^{{\rm obs}}_e | \mu, \sigma)\,.
	\end{align}
	\end{subequations}%
	For our uniform sampling prior, the $W(\btheta)$ term in Eq.~\eqref{eq:the-generic-integral} is also uniform and can be dropped.
	
	Since we assume that we only have access to draws from $\phi_{[0,1]}(\chi | \chi^{{\rm obs}}_e, \sigmanoise)$ without knowledge of its true functional form, we cannot evaluate Eq.~\eqref{eq:ToyExampleIntegral} exactly and must instead approximate $\mathcal{L}(\chi^{{\rm obs}}_e | \mu, \sigma)$ from these samples. 
	
	\begin{figure*}[t]
		\begin{center}
			{\includegraphics[width=0.8\paperwidth]{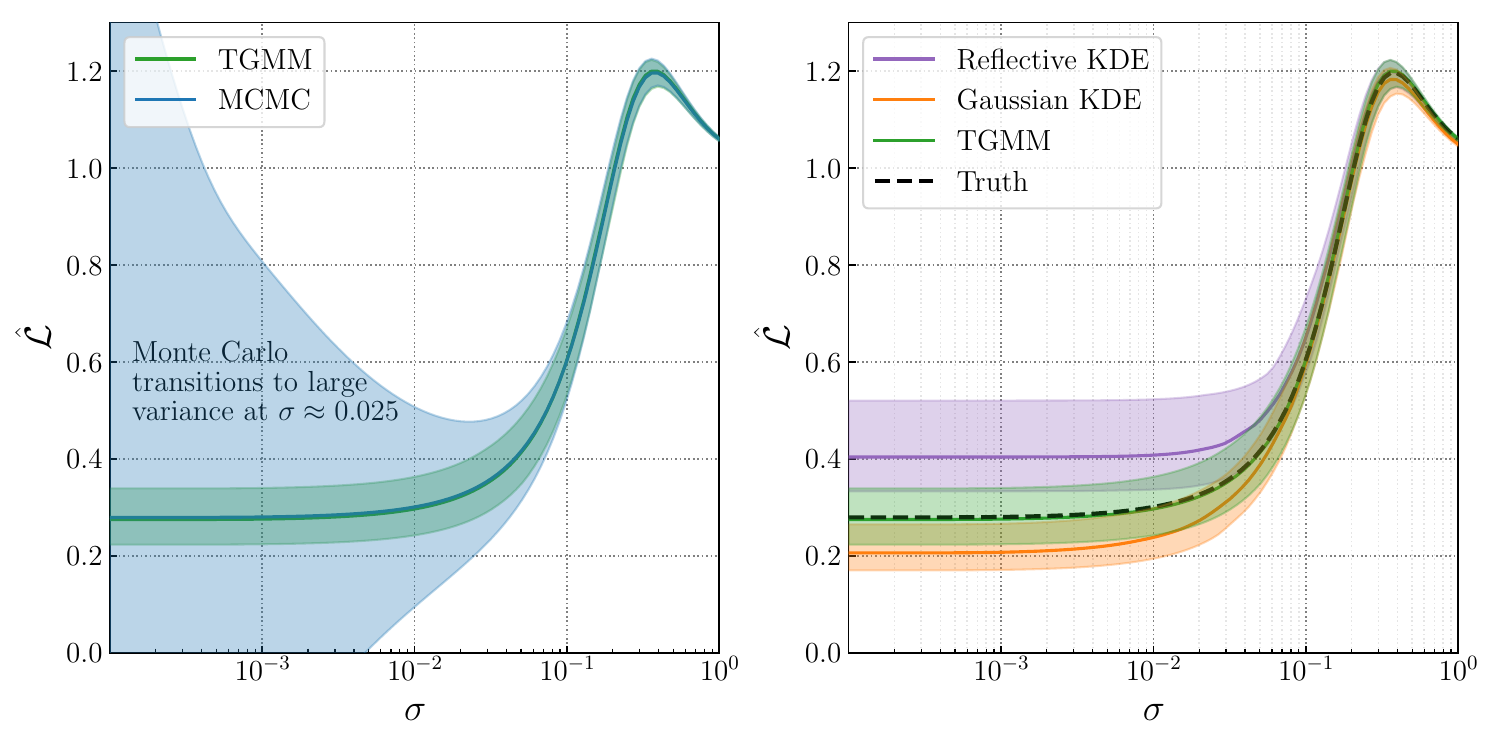}}
		\end{center}
		\vspace*{-5mm}
		\caption{Bias and variance comparisons between different estimation methods in a toy example. Both panels show the same comparison: estimating the likelihood for a posterior with shape $\phi_{[0,1]}(0.4, 0.2)$ and a population model that gets increasingly compact near the edge $\phi_{[0,1]}(0, \sigma), \sigma \to 0$. 
		The variance of the likelihood estimate (95\% confidence interval shown here) can blow up for values of the population deviation parameter smaller than $\sigma \approx 0.025$ for 1000 samples. 
			The confidence intervals of the \gls{TGMM} (using bootstrapped estimates) are much tighter and around the correct value. 
			Additionally \gls{KDE} techniques can fix this blow-up in the variance estimate, but can give biased estimates as we approach the edge. 
			This efficiency and bias only gets worse with an increase in the number of dimensions.}
		\label{fig:IllustrativeExample}
	\end{figure*}
	
	We show how the different estimation techniques for $\mathcal{L}(\chi^{{\rm obs}}_e \mid \mu, \sigma)$ above compare in terms of their variance and bias (Fig.~\ref{fig:IllustrativeExample}). 
	We consider the specific case of our toy population model evaluated at the boundary $\mu = 0$ and look at the evolution of the estimators as $\sigma \to 0$. 
	We consider a particular event likelihood that follows the shape of a truncated normal with parameters $\phi_{[0,1]}(0.4, 0.2)$, which corresponds to a detected value of $\chi^{{\rm obs}}_e = 0.4$ and detection noise  $\sigma_{\text{noise}} = 0.2$.  
	We subsequently take 1000 draws from this likelihood and utilize the estimator to get one estimate of the integral; for the \gls{MC} estimator, we compute the bias and variance analytically (Appendix~\ref{sec:MCVariance}).
	
	
	To begin with, the \gls{MC} estimate transitions to a regime of very large variance at $\sigma \approx 0.025$. 
	This transition is expected to happen at an increasingly larger value of $\sigma$ as the dimensionality of the problem increases \citep{Talbot2023}.
	Although the \gls{MC} estimates are unbiased (as can be seen by the dotted blue line in the second panel of Fig.~\ref{fig:IllustrativeExample}), the large variance makes it unusable for practical purposes.
	On the other hand, both types of \glspl{KDE} retain a relatively small variance throughout the domain. However, they fail to converge to the correct value of the likelihood estimate at the edge. 
	
	Our toy example highlights key limitations of commonly employed integration methods for astrophysical population analysis near parameter boundaries: standard \gls{MC} approaches suffer severe variance limitations, while \glspl{KDE} alleviate variance at the expense of systematic boundary biases---this is the problem we set out to address.
	We will next introduce our \gls{TGMM} method, which faithfully captures the true value of the likelihood $\mathcal{L}(\chi^{{\rm obs}}_e \mid \mu, \sigma)$ with minimal bias, all while having a controlled variance near the edge and computational advantage in this particular test.

	
	\section{Our proposed method: Truncated Gaussian Mixture models }
	
	Section~\ref{sec:GeneralProcedure} showed that, when
	the target population is sharply localized near a parameter
	boundary, existing estimators incur either large variance or large boundary bias.  
	Our \gls{TGMM} method addresses this by retaining the analytic convenience of a kernel approach without surrendering boundary fidelity.
	
	In what follows, we denote the set of parameters as $\btheta$ and let their boundaries be specified by hyper-corner vectors $\ba$ and $\bb$, where $\ba$ sets the lower limit and $\bb$ sets the upper limit for all bounded dimensions.
	For parameters that are unbounded, their entries in $\ba$ or $\bb$ can be taken to infinity to remove one of the boundaries. 
	
The hybrid \gls{KDE} methods described in Section.~\ref{sec:techniques-for-computing} factorize the parameter space into dimensions with an analytically integrable population likelihood and others where we fall back on \gls{MC} estimation. In the same spirit, we separate our parameters $\btheta$ into two sectors based on whether the per-event integral of Eq.~\eqref{eq:the-generic-integral} admits analytic or semi-analytic evaluation under the chosen kernel. The Analytic Sector ($\btheta^a$) consists of parameters for which this integral can be computed in closed or semi-closed form for the chosen pair of population model $p(\btheta\mid\tilde{\bLambda})$ and kernel $\phi$; truncated Gaussians, uniform distributions, and Heaviside functions are examples that yield tractable integrals. The remaining parameters form the Sampled Sector ($\btheta^s$), where we estimate the integral via standard Monte Carlo techniques.
	While this breakup offers increased flexibility, it does impose the constraint that the population model itself separate into a product of these sectors, i.e., the population model must not correlate $\btheta^a$ and $\btheta^s$.
	Formally, that means the model must factorize as
	\begin{equation}
		\label{eq:separate_single_pop_model}
		p(\btheta \mid \Lambda) = p(\btheta^a \mid \Lambda^a)\, p(\btheta^s \mid \Lambda^s) \,,
	\end{equation}
	where $\btheta = \btheta^a\oplus\btheta^s$ and $\bLambda = \bLambda^a\oplus\bLambda^s$. 
	Mixtures of such product distributions could also be allowed, i.e.,
	\begin{equation}
		p(\btheta \mid \Lambda) = \sum_k p(\btheta^a_k\mid \Lambda^a_k)\, p(\btheta^s_k \mid \Lambda^s_k)\,,
	\end{equation}
	for $k$ components.
	While it is straightforward to generalize to the mixture case, we describe our method using only one term in the population model as in Eq.~\eqref{eq:separate_single_pop_model}.
	
Crucially, in the Analytic Sector we would like to utilize analytic estimates of the posterior over $\btheta^a$ using a \textit{truncated-Gaussian mixture} (TGMM) -- setting our choice of kernel to be Truncated Gaussian, while setting our population model to be also a \gls{TGMM}. 
This pairing of a truncated Gaussian kernel with truncated Gaussian population components yields closed-form expressions for the per-event integrals like Eq.~\eqref{eq:KDE-integral}, enabling accurate boundary behavior and fast evaluation. 
In the following, we fix our population model to be a mixture of truncated Gaussians in the Analytic Sector to capture edge behavior of bounded parameters.
	Namely,
	\begin{equation}
		\begin{aligned}
			p(\btheta \mid \bLambda) &= p(\btheta^s \mid \bLambda^s)\sum_l^L \eta_l\, \phi_{[\ba, \bb]}(\btheta^a | \bmu_l, \bSigma_l) \,,
		\end{aligned}
		\label{eq:modelrestrictionextra}
	\end{equation}
	where $L$ is the number of components in the population model and $\eta_l$ are the mixture weights (with $\sum_l \eta_l = 1$).
	The formalism is the same for any other choice as long as it is analytically integrable against the \gls{TGMM} kernel.
	
	With this factorized population model,
	we can revisit the integral of Eq.~\eqref{eq:the-generic-integral},
	\begin{equation} \label{eq:tgmm-integral-derivation-first-step}
		\begin{aligned}
			I(\bLambda)
			=  \int & \frac{p(\btheta^s \mid \bLambda^s)}{W^s(\btheta^s)}\left[\sum_l^L \eta_l \, \phi_{[\ba, \bb]}(\btheta^a | \bmu_l, \bSigma_l)\right]
		\\ & \times
		 \frac{p(\btheta)}{W^a(\btheta^a)} \d\btheta\, ,
		\end{aligned}
	\end{equation} 
	where we have assumed that the weight function $W(\btheta)$ can separate into $W^s(\btheta^s)W^a(\btheta^a)$.
	When this separation cannot be achieved, one can fit the whole dataset with weights $W_i = W(\btheta_i)$ during the fitting procedure, as opposed to just the analytic portion. We have found that this degrades the fits slightly if the weights span a large dynamic range.
	
	We then fit the weighted probability distribution in the last factor of Eq.~\eqref{eq:tgmm-integral-derivation-first-step} to a \gls{TGMM} (Appendix \ref{sec:TGMM}).
	In doing so, we also impose that the individual Gaussian correlation matrices separate across the analytic and sampled sectors,
	 i.e., $\bSigma_k = \bSigma^a_k \oplus \bSigma^s_k$.
	 This gives
	\begin{equation} \label{eq:TGMM-posterior-fit}
		\begin{aligned}
			\frac{p(\btheta)}{W^a(\btheta^a)} &\approx \sum_k^K w_k \phi_{[\boldsymbol{a},\boldsymbol{b}]}(\btheta \mid \boldsymbol{\mu}_k ,\boldsymbol{\Sigma}_k) \\ &=  \sum_k^K w_k \phi_{[\boldsymbol{a},\boldsymbol{b}]}(\btheta^a \mid \boldsymbol{\mu}^a_k ,\boldsymbol{\Sigma}^a_k) \phi_{[\boldsymbol{a},\boldsymbol{b}]}(\btheta^s \mid \boldsymbol{\mu}^s_k ,\boldsymbol{\Sigma}^s_k)\,.
		\end{aligned}
	\end{equation}
	where $K$ is the number of components in the TGMM fit and $w_k$ are the mixture weights (with $\sum_k w_k = 1$).
	While this factorization constraint is a limitation, one can always add enough TGMM components to effectively capture correlations in the likelihoods across sectors.
	
Above, we use \glspl{TGMM} to fit the target density $p(\btheta) / W^a(\btheta^a)$ rather than $p(\btheta)$ itself. 
This ensures that the analytic part of the integrand matches the truncated-Gaussian kernel, since the ability to analytically integrate applies to pairs of densities (population and kernel) but not to arbitrary weight functions.
	This allows us to further rewrite the integral of Eq.~\eqref{eq:tgmm-integral-derivation-first-step} as
	\begin{align} \label{eq:IntegralTGMMDerivation}
		I(\bLambda) &=
		 \sum_{kl}^{K,L} \eta_l w_k \int \frac{p(\btheta^s \mid \bLambda^s)}{W^s(\btheta^s)}  \phi_{[\boldsymbol{a},\boldsymbol{b}]}(\btheta^s \mid \boldsymbol{\mu}^s_k ,\boldsymbol{\Sigma}^s_k)\, \d\btheta^s \nonumber \\ 
		&\times \int \phi_{[\ba, \bb]}(\btheta^a | \bmu_l, \bSigma_l)\, \phi_{[\boldsymbol{a},\boldsymbol{b}]}(\btheta^a \mid \boldsymbol{\mu}^a_k ,\boldsymbol{\Sigma}^a_k)\, \d\btheta^a \\
		&= \sum_{kl}^{K,L} \eta_l w_k \left\langle   \frac{p(\btheta^s|\Lambda^s)}{W^s(\btheta^s)}  \right\rangle_{\btheta \sim \mathcal{N}_{[\boldsymbol{a},\boldsymbol{b}]}(\boldsymbol{\mu}^s_k ,\boldsymbol{\Sigma}^s_k)} \nonumber\\ 
		&\times F_{[\ba, \bb]}(\bmu^a_k, \bSigma^a_k,\bmu_l, \bSigma_l)\,,
	\end{align}
	where $F_{[\ba, \bb]}(\bmu_1, \bSigma_1, \bmu_2, \bSigma_2)$ is the integral of the product of two truncated multivariate gaussians with parameters $(\bmu_1, \bSigma_1)$ and $(\bmu_2, \bSigma_2)$ over the hypercube with bounding corners $\ba$ and $\bb$.
	While we have assumed that the truncation domain in the population model $[\ba, \bb]$ is the same as that in our \gls{TGMM} fits, one can relax that constraint, implying that the function $F$ in Eq.~\eqref{eq:IntegralTGMMDerivation} depends on truncation regions $[\ba_1, \bb_1]$ from the \gls{TGMM} fits and $[\ba_2, \bb_2]$ from the population model. In general the expression for $F$ can be found in Appendix~\ref{sec:different-bounded-corners}.
	
	Once we have evaluated $F_{[\ba, \bb]}(\bmu^a_k, \bSigma^a_k,\bmu_l, \bSigma_l)$, we could then evaluate the expectation over the sampled parameters using \gls{MC} sampling,
	\begin{equation}
		\label{eq:SampledExpectation}
		\left\langle   \frac{p(\btheta^s|\Lambda^s)}{W^s(\btheta^s)}  \right\rangle_{\btheta \sim \mathcal{N}_{[\boldsymbol{a},\boldsymbol{b}]}(\boldsymbol{\mu}^s_k ,\boldsymbol{\Sigma}^s_k)} \approx \frac{1}{N}\sum_{n=1}^{N} \frac{p(\btheta^s_n \mid \bLambda^s)}{W^s(\btheta_n)}\,,
	\end{equation}
	where we have averaged over $N$ samples from the $k$th Gaussian component, $\mathcal{N}_{[\boldsymbol{a},\boldsymbol{b}]}(\boldsymbol{\mu}^s_k ,\boldsymbol{\Sigma}^s_k)$.

	However, we have found better estimates of $I(\bLambda)$ if we do not generate new samples from $\mathcal{N}_{[\boldsymbol{a},\boldsymbol{b}]}(\boldsymbol{\mu}^s_k ,\boldsymbol{\Sigma}^s_k)$,
	but instead use the original event posterior samples for these estimates. 
	This approach can help address some limitations of the \gls{TGMM} as a density estimator. 
	For instance, the \gls{TGMM} may struggle to capture non-trivial discontinuities present in the original posterior distribution. By utilizing the original posterior samples, the estimator becomes more robust to such features.

	The TGMM fitting procedure provides, for each sample, the probability that it belongs to each mixture component (a ``soft assignment"). 
	We assign each sample to a specific component by randomly choosing according to these probabilities, resulting in $k$ groups of data points, each of size $N_k$ (with $N = \sum_k N_k$). 
	With this grouping, we then perform the Monte Carlo sum in Eq.~\eqref{eq:SampledExpectation} using the original samples.
	\begin{equation}
	\begin{aligned}
		I(\bLambda) 
		\approx   \sum_{kl} & \eta_l w_k 
		\left[ \frac{1}{N_k}\sum_{i=1}^{N_k} \frac{p(\btheta^s_i|\Lambda^s)}{W^s(\btheta^s_i)}  \right] 
		\notag \\ \times &
		F_{[\ba, \bb]}(\bmu^a_k, \bSigma^a_k,\bmu_l, \bSigma_l)\,.
		\label{eq:hybrid-scheme}
	\end{aligned}
	\end{equation}
	This yields the final expression for the integrals required when evaluating the event-level likelihood $\mathcal L(d_e | \bLambda)$ and detection efficiency $\xi(\bLambda)$ during population inference.	

	\begin{table*}
		\centering
		\caption{Summary of models and prior ranges for Section~\ref{sec:GWApplication}}
		\label{tab:WidePriorModels}
		\begin{tabular}{llcc}
			\toprule
			\textbf{Sector} & \textbf{Model} & \textbf{Reference} & \textbf{Priors}\\ 
			\toprule
			\midrule
			Mass & \texttt{Power Law + Peak} & \cite{LVKPop}  (B4)& Same as \cite{LVKPop}\\ 
			Spin Orientation & \texttt{DEFAULT} & \cite{LVKPop} (B20) & Same as \cite{LVKPop}\\ 
			Spin Magnitude & Truncated Gaussian & $\sim \mathcal{N}_{[0,1]}(\mu_\chi, \sigma_\chi) $ & 
			$\mu_\chi \sim U(0,1), \sigma_\chi \sim U(0.15, 1)$\\ 
			Redshift & \texttt{PowerLaw} & $ \propto \frac{dV_c}{dz} (1+z)^{\kappa -1 }$ & Same as \cite{LVKPop}\\ 
			\bottomrule
		\end{tabular}
	\end{table*}
	
	Equation \eqref{eq:IntegralTGMMDerivation} represents a significant simplification: armed with \gls{TGMM} fits to the data, we can sample the components for the \gls{MC} integral over the ``sampled'' population model, while analytically integrating over the bounded sector.
	This is advantageous for both efficiency and accuracy,
	as we demonstrate next in the context of \gls{BH} spin measurements with \glspl{GW}.

	Further details regarding the fitting of \glspl{TGMM} to data can be found in the appendices.
	More specifically, Appendix~\ref{sec:TGMM} discusses the expectation-maximization algorithm and modifications for fitting truncated Gaussian mixtures. 
	Appendix~\ref{sec:BoundaryKDE} discusses the derivation and use of a boundary-corrected \gls{KDE} that improves fitting performance of the \gls{TGMM} fit. 
	Lastly, Appendix~\ref{sec:different-bounded-corners} provides the general analytic expression for the integral of a product of two truncated multivariate Gaussians and Appendix~\ref{sec:ComputationOfIntegral} discusses computational aspects involved in the numerical evaluation of that analytic expression.

	\section{Implementation for gravitational waves}
	\label{sec:GWApplication}
	In this section we reproduce the results of the fiducial population analysis presented in \cite{LVKPop}, 
	while applying a small modification to the model for the spin distributions to adapt them to our methods.
	We carry out this inference using both \gls{MC} methods and our \gls{TGMM} approach.
	We find close agreement between these two approaches in regions where both are expected to be accurate. 
	This validates our methods in a realistic setting while including selection effects and modelling both the \gls{MC} and analytic sectors.
	In addition, we compare the variance in our estimates of the event-level margnalized population likelihood and detection efficiency with that of standard \gls{MC} methods, and explore wider regions of hyper-parameter space than the analysis of \cite{LVKPop}.
	
	
	\subsection{Gravitational wave data and population model}
	\label{sec:Data}
	
	We follow the event-selection protocol utilized by \cite{LVKPop}, and select 69 \gls{BBH} events that pass a FAR threshold of $< 1 \text{yr}^{-1}$. 
	For these events we use the posterior samples parameter estimation performed using the \texttt{IMRPhenomXPHM} waveform \citep{Pratten:2020ceb} and presented with from GWTC-2.1 \citep{LIGOScientific:2021usb} and GWTC-3 \citep{KAGRA:2021vkt}. 
	These samples are available as open data~\citep{LIGOScientific:2019lzm,KAGRA:2023pio} at \cite{GWOSC}. 
	To quantify the selection effects, we use the sensitivity estimates described in \cite{LVKPop} and provided by the LVK~\cite{LIGO-SearchSensitivity, essick_semianalytic_2023}. 

	For our population model, we mimic the fiducial model of \cite{LVKPop}, with the notable exception that we utilize a truncated normal rather than a
	Beta distribution to model the population from which the of spin magnitudes $\chi_i$ are independently drawn.
	The elements of our model and our choice of hyperpriors is given in Table~\ref{tab:WidePriorModels}.
	
	\subsection{Fitting \glspl{TGMM} to gravitational wave data}
	
	
	We next describe the \gls{TGMM} fit hyperparameters used to get a good fit to the posterior samples.
	Both the posterior samples and sensitivity estimates constitute weighted datasets, over which we take an expectation of the population model. 
	The posterior samples in such an analysis are reported with the prior used fror sampling $W(\btheta)= 1/\pi(\btheta)$, and the sensitivity estimates are reported with the draw probability used to generate the injections $W(\btheta)= 1/p_{\text{draw}}(\btheta)$ which plays an analagous role to the prior mentioned previously, as seen in Eq.~\eqref{eq:detectionefficiency}.
	
	We  analyze the spin magnitudes $\chi_1, \chi_2$ and the spin orientations $\cos\theta_1, \cos\theta_2$ analytically, while falling back on \gls{MC} estimates for the primary mass, mass ratio and redshift parameters ($m_1, q, z$). 
	For the posterior samples and the selection injections, we utilize the fact that the priors and draw probabilities are flat in the space of spin magnitude and spin orientation and so we set the analytic portions of these weights to unity ($W^a(\btheta^{a}) = 1$).
	Additionally, we use the reported weights in the remaining dimensions augmented with appropriate Jacobian factors as described in \cite{LVKPop}. 
	Performing these steps gives us the weighted datasets of the event posteriors and the selection injection sets.

	\begin{figure*}
		\label{fig:EventFitExample}
		\begin{center}		
			\includegraphics[width=0.8\paperwidth]{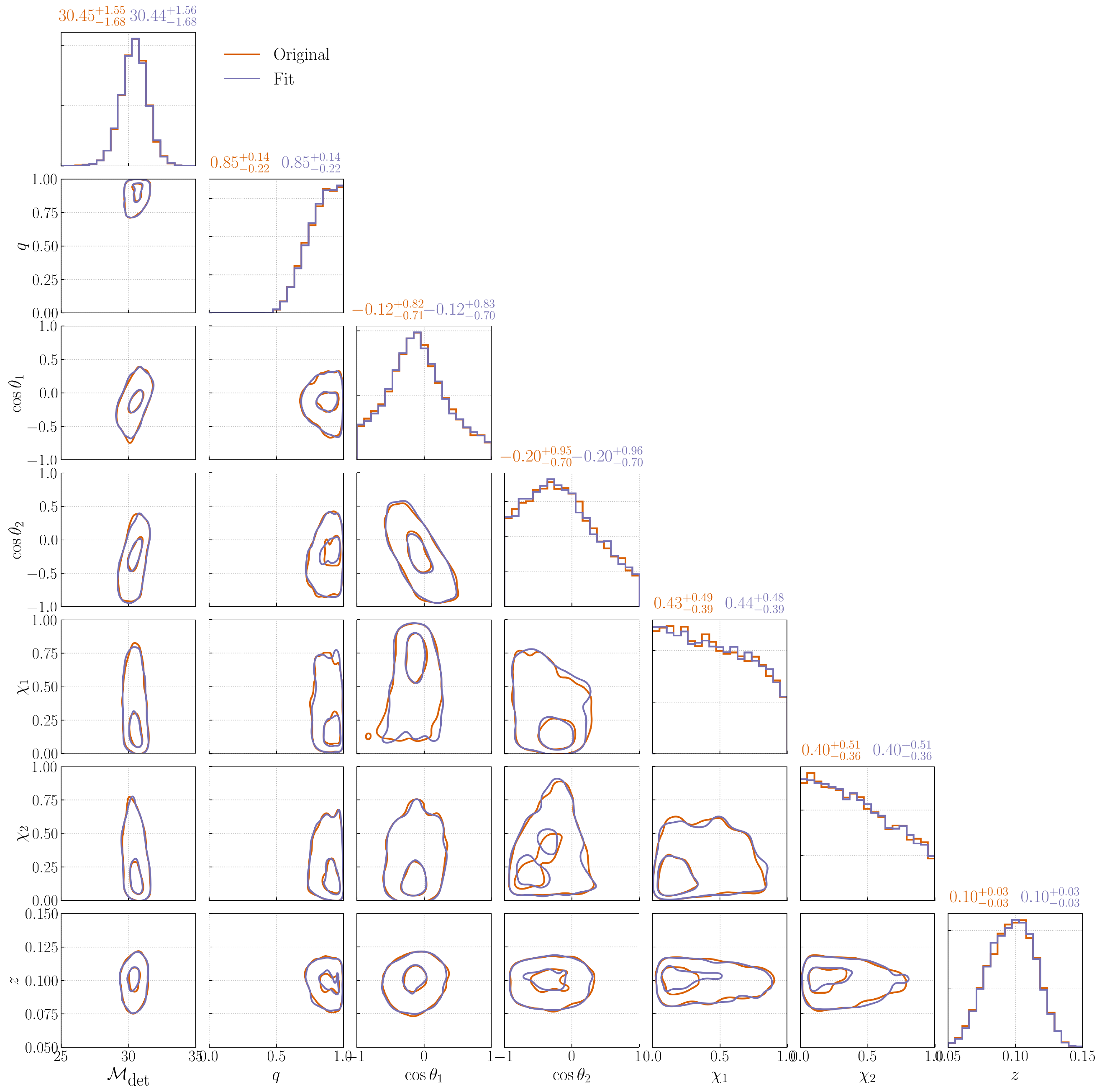}
		\end{center}
		\caption{Corner plot comparing the \gls{TGMM} fit and the posterior samples for GW150914.}
		\label{fig:TGMMCompare}
	\end{figure*}
	
	Subsequently, we fit \glspl{TGMM} to these datasets (see appendices for details).
	For the event level posteriors we fit $K = 15$ components with the covariance matrix consisting of three blocks: a mass and redshift sector, a spin magnitudes sector and a spin orientations sector. 
	Each block can harbor correlations within the block but there can be no correlations between parameters of different blocks. 
	We use the boundary unbiasing-technique 
	of Appendix \ref{sec:BoundaryUnbiasing}
	for $N_{\text{KDE}} = 100$ iterations and then subsequently let the algorithm run for an additional $200$ iterations. 
	For a better fit, we transform the primary source mass variable to the detector chirp mass, and fit the posterior samples in this new latent space. 
	Once we have the resulting assignments of original samples to components, along with the \gls{TGMM} parameters, we can then transform back to the primary source mass variable, and gather the assigned samples for every component. 
	We can then use Eq.~\eqref{eq:hybrid-scheme} to compute the population likelihood for a given event. 
	
	\begin{figure*}[tb]
		\begin{center}
			\includegraphics[width=0.4\paperwidth]{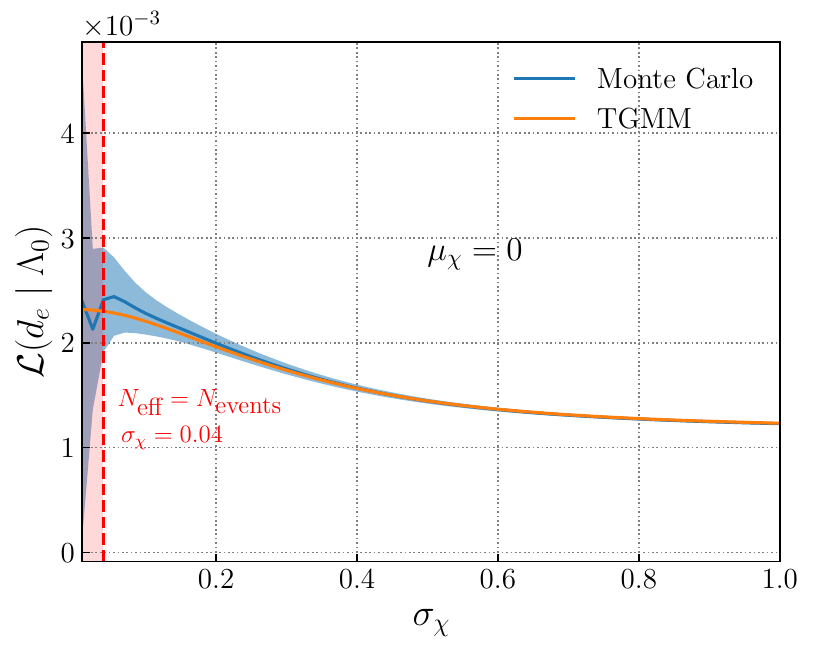}
			\includegraphics[width=0.4\paperwidth]{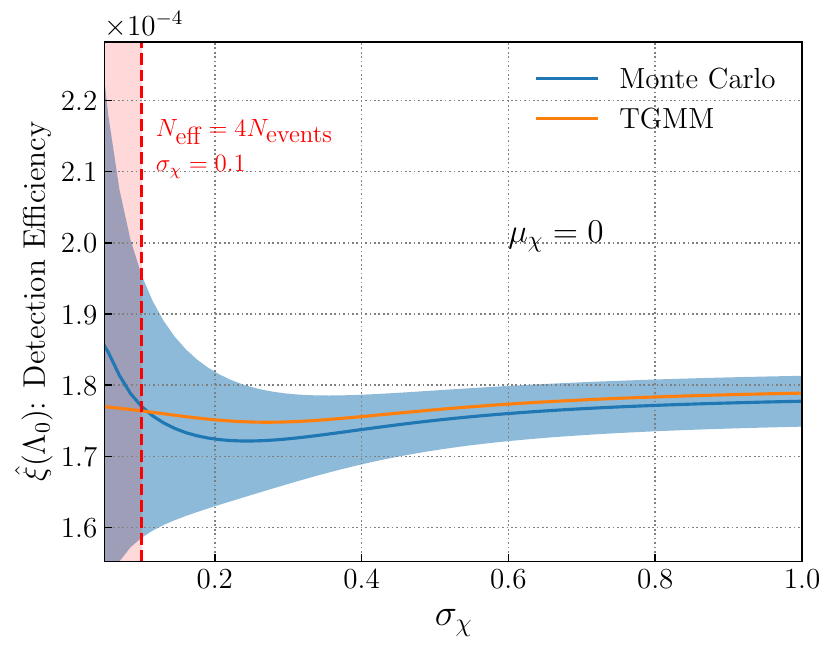}
		\end{center}
		\caption{Comparison of integral estimates from both the \gls{TGMM} method and \gls{MC} when evaluated at the median hyper-parameters ($\bLambda_0$) of our population model, but with $\mu_\chi = 0$ and $\sigma_\chi$ allowed to vary. The $2\sigma$ variance of the \gls{MC} estimate is shown as blue bands. 
			In both cases the \gls{TGMM} estimate remains consistent with the \gls{MC} estimate as the width of the spin magnitude population narrows.
			{\it Left:}  Comparison of the marginalized likelihoods for GW150914.
			{\it Right:} Comparison of the detection efficiency.
		}
		\label{fig:IntegralVaryWithSigma}
	\end{figure*}

	For the selection injections we use $K = 40$ components, but turn off any correlations between parameters (i.e., we use diagonal covariances). 
	We then utilize the boundary-unbiasing technique  
	for $N_{\text{KDE}} = 100$ iterations and then subsequently let the algorithm run for $200$ additional iterations. 
	Additionally we transform the samples in the sampled parameters using the transformation 
	$(m_1, q) \to (\log(m_1/M_\odot - 2), (q m_1 - 2 M_\odot)/(m_1 - 2 M_\odot))$, 
	which allows for a better resolution of the $m_1, m_2 > 2 M_\odot$ cut inherent in the injection set, and then subsequently fit the \gls{TGMM} in the transformed space. 
	
	Figure~\ref{fig:TGMMCompare} demonstrates an example of the agreement between our \gls{TGMM} fits and the samples we fit over, for the event GW150914.
	While the posteriors for each event and  $\pdet(\btheta | \bLambda_\emptyset)$ with $W^a(\btheta) = 1$ display good visual agreement with the weighted samples we fit over, this is not sufficient to guarantee that the integrals $I(\bLambda)$ we use them to compute are accurate.
	To demonstrate the accuracy of our estimate of $I(\bLambda)$ we turn to a detailed comparison of our integral estimates, and our final comparison of population inferences between the methods.
	
%


	\subsection{Validation of Fits}
	
	In this section we validate the effectiveness of the \gls{TGMM} fit
	by comparing estimates of the integrals involved in the population inference likelihood between our \gls{TGMM} and \gls{MC} methods.
	In particular, we compare evaluations of the marginalized likelihood $\mathcal L(d_e | \bLambda_0)$  for the GW150914 event, as well as the overall detection efficiency $\xi(\bLambda_0)$.
	For these comparisons we use the median hyperparameters $\bLambda_0$ from our full population inference presented below in Section~\ref{sec:WidePriorEquivalence}, except that for the spin distribution we take $\mu_\chi =0$ and vary the width parameter $\sigma_\chi$ to explore the two approaches at the edge of hyperparameter space.

	Figure~\ref{fig:IntegralVaryWithSigma} illustrates these estimates these results, along with the $2\sigma$ variance of the \gls{MC} method. 
	We see that the \gls{TGMM} results are consistent within these error bands.
	Meanwhile, the variance of the \gls{MC} estimates balloons in the limit $\sigma_\chi \to 0$ as expected.
	One method to track the failiure of the \gls{MC} estimate is to track the effective number of samples that inform the estimate $N_{\text{eff}}$ (see \cite{Farr:2019rap, LVKPop} for details).
	Additionally we mark the $\sigma_\chi$ value where the effective number of samples $N_{\text{eff}}$ used in the \gls{MC} sum falls below a multiple of the number of events $N_{\text{events}}$ , beyond which the estimates are not expected to be sufficiently accurate for the analysis~\citep{Farr:2019rap}.
	For the $\mathcal L(d_e | \bLambda)$,  $N_{\text{eff}} > N_{\text{events}}$ is used as a cut on the population posteriors in \cite{LVKPop}, while for $\xi(\bLambda)$, $N_{\text{eff}} > 4N_{\text{events}}$ is required following that reference.
	
	\begin{figure*}[tb]
		\begin{center}
			\includegraphics[width=0.4\paperwidth]{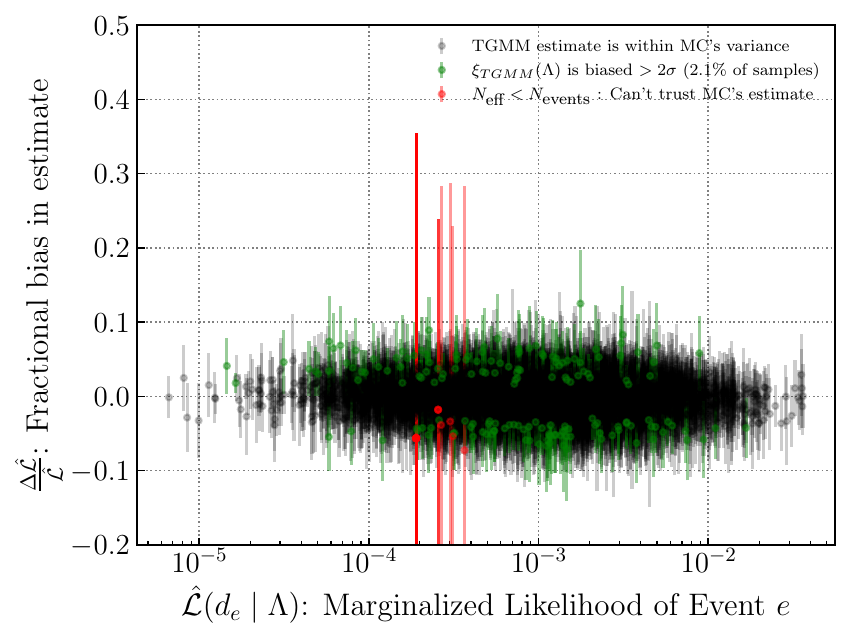}
			\includegraphics[width=0.4\paperwidth]{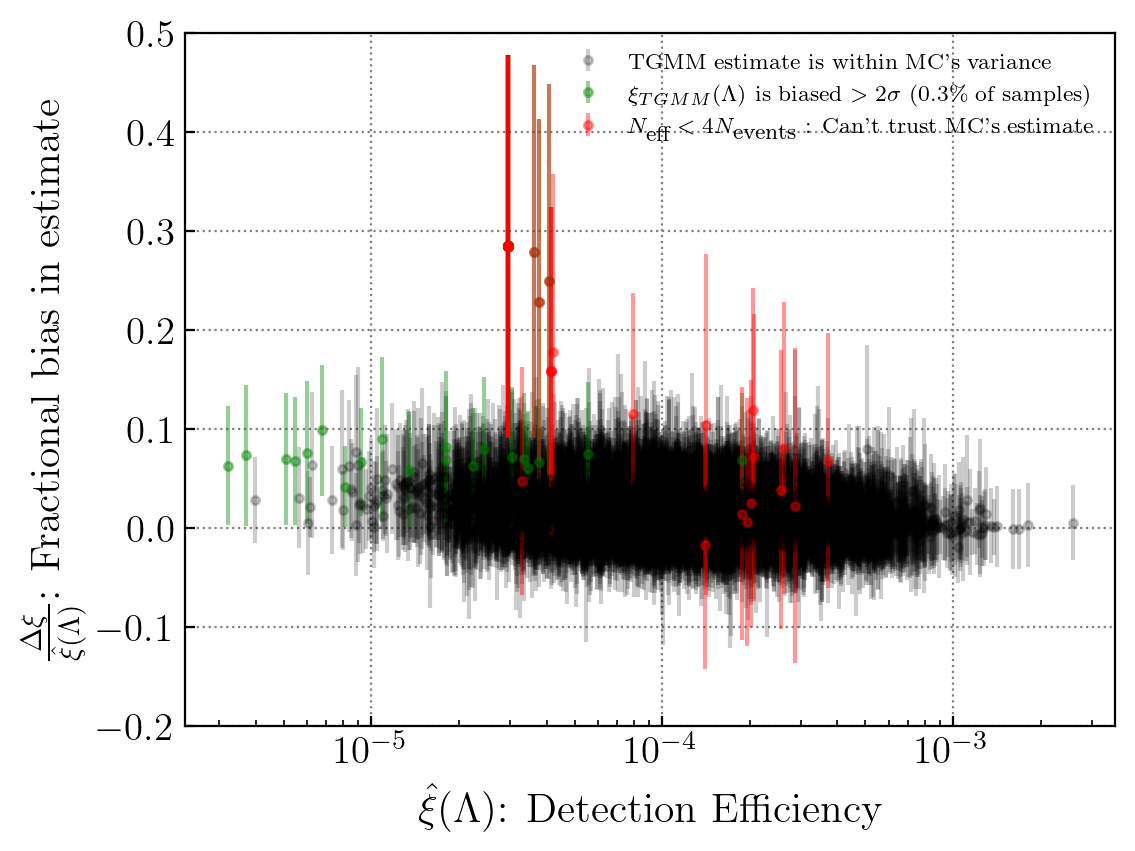}
		\end{center}
		\caption{The fractional bias  between the between \gls{TGMM} and \gls{MC} integrals, relative to the \gls{TGMM} result. 
		Each point is the integral estimate evaluated at a  hyper-posterior sample $\bLambda_i$ from our full population inference using the model described in Table~\ref{tab:WidePriorModels}. 
		The error bars encapsulate the $2 \sigma$ deviation of the estimate as expected by the variance of the \gls{MC} estimator.
		The error bars denote the $2 \sigma$ deviation of the estimate as expected by the variance of the \gls{MC} estimator.
		We color those samples outside the $2 \sigma$ range green, and those which have $N_{\text{eff}}$ too small red.
		{\it Left:} The fractional bias in the marginalized likelihood estimate for GW150914. 
		We find that only $\approx 2.1\%$ of samples are outside the $2 \sigma$ range.
		{\it Right:} The fractional bias in the detection efficiency. 
		We find that only $\approx 0.3\%$ of samples are outside this $2 \sigma$ range.}
		\label{fig:FitVariance}
	\end{figure*}

	\begin{figure}[t]
		\begin{center}
			\includegraphics[width=0.4\paperwidth]{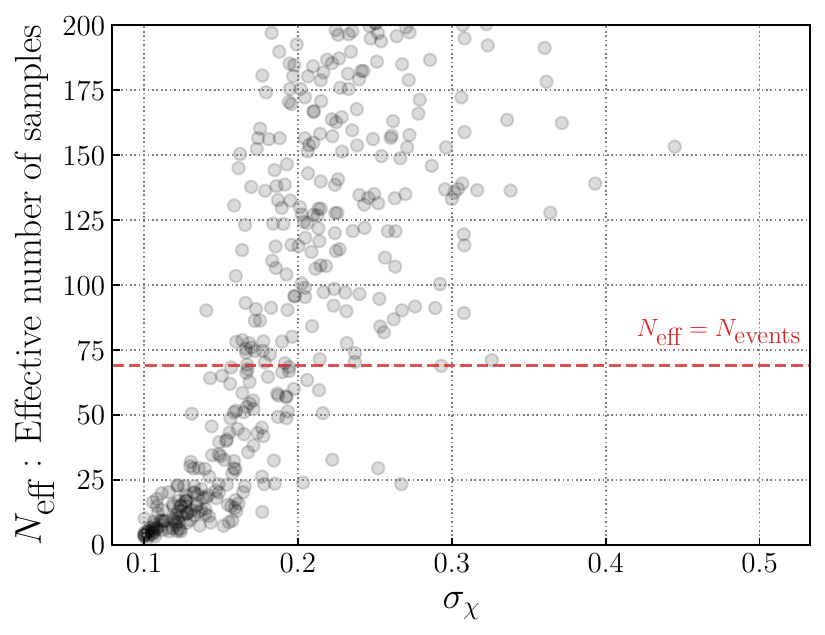}
		\end{center}
		\caption{For each hyper-posterior sample we compute the number of effective samples in the \gls{MC} estimate for each event. 
		Since our likelihood estimate is only valid if we have good convergence for all events we plot the smallest $N_{\text{eff}}$ for each sample on the $y$-axis (this is usually GW190517). 
		We can see that for $\sigma_\chi < 0.15$ we are in a regime where the \gls{MC} estimate has too few points. 
		As such we only use posterior samples with $\sigma_\chi > 0.15$ to directly compare the \gls{TGMM} and \gls{MC} results.}
		\label{fig:N_eff_tracker}
	\end{figure}

	As a further comparison, we compute the fractional difference between the two integral estimation methods across all hyper-posterior samples from our fiducial population inference.
	These are given in Figure~\ref{fig:FitVariance}, again for the marginalized likelihood of GW150914 and for the detection efficiency.
	Our results indicate agreement with the \gls{MC} everywhere where that method is expected to be accurate, with significant deviations occurring only where $N_{\text{eff}}$ is too small for the given sample $\bLambda_i$.

	\subsection{Validation against existing population inference for GWTC-3}
	\label{sec:WidePriorEquivalence}
	Having seen that our \gls{TGMM} fitting yields reliable estimates for the ingredients in the population inference, we compare the results of a full population analysis (modeling the mass, spin and redshift parameters simultaneously) mimicking that of \cite{LVKPop}.
	We carry our inference using our \gls{TGMM} method and compare to the same analysis using the standard \gls{MC} approach, with $10^5$ samples per event.
	The resulting hyperposterior samples are those used above in Figure~\ref{fig:FitVariance}.
	We do not expect our two results to be in precise agreement, since there are parameter regions where the \gls{MC} estimates are unreliable.
	In order to validate our method, we find the region where the \gls{MC} estimates are reliable using a criterion based on $N_{\text{eff}}$.
	
\begin{figure}[h]
		\begin{center}
			\includegraphics[width=0.4\paperwidth]{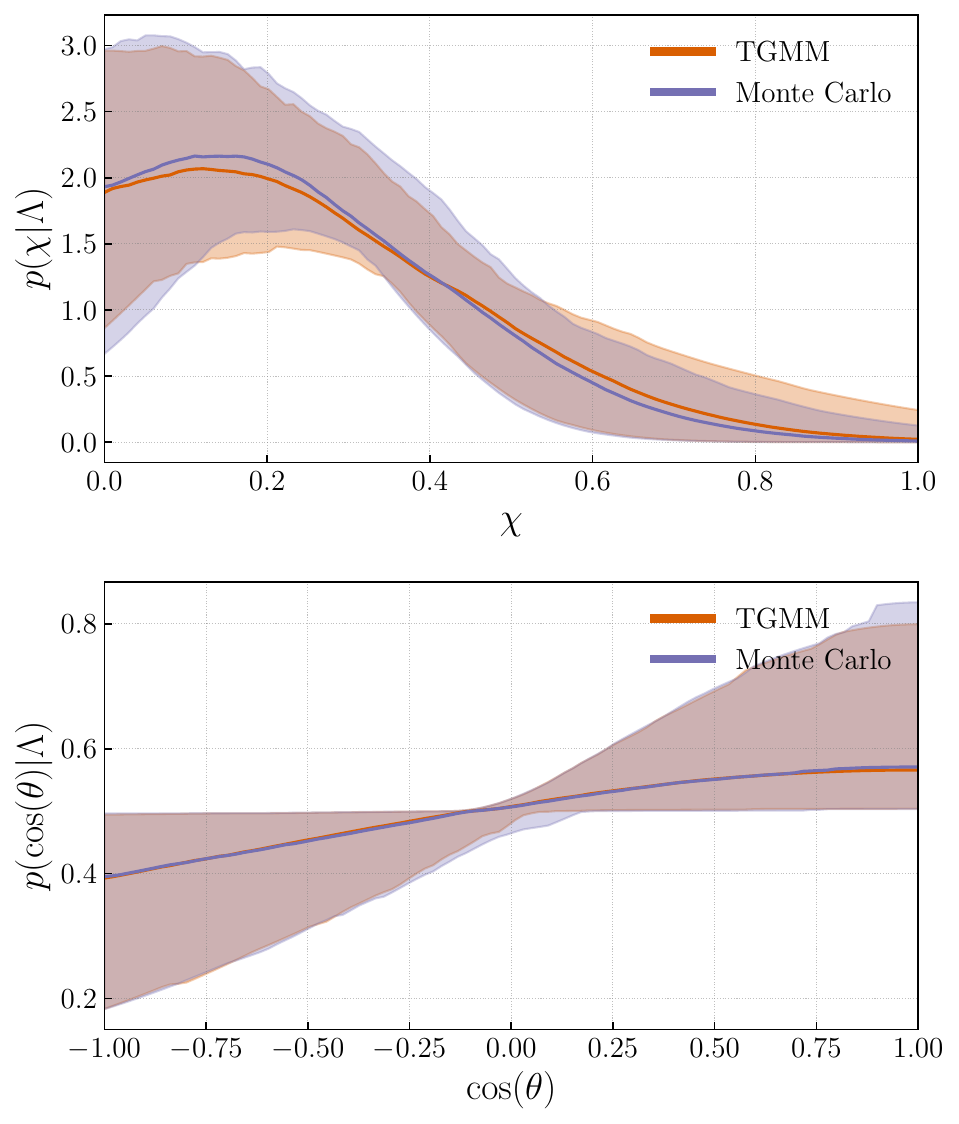}
		\end{center}
		\caption{Comparison between the posterior estimates using the importance sampling techniques and the \gls{TGMM} method with a prior cut at $\sigma_\chi > 0.15$.}
		\label{fig:SampledComparison}
	\end{figure}

	Figure~\ref{fig:N_eff_tracker} plots the smallest $N_{\text{eff}}$ over all events against the width parameter of the spin magnitude model, $\sigma_\chi$.
	From this we see that $N_{\text{eff}} > N_{\text{events}}$ if we require $\sigma_\chi > 0.15$, a cut which we adopt for direct comparison of our results.
	In Figure~\ref{fig:SampledComparison} we show the posterior predictive distributions for spin magnitudes and tilt angles between the two methods.
	The two methods agree quite well in regimes where the usual importance sampling technique is trustworthy. 
	We provide further comparison plots in Appendix~\ref{sec:AdditionalResults}.

	\section{Zero-spin sub-populations of \glspl{BBH} from the GWTC-3 Catalog}
	\label{sec:ZeroSpin}
	
	\begin{figure}
		\begin{center}
			\includegraphics[width=0.4\paperwidth]{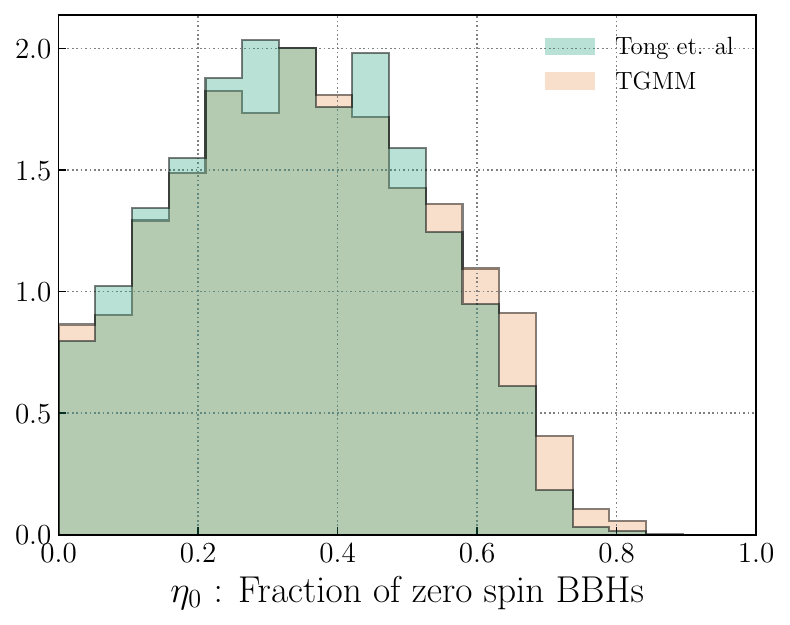}
		\end{center}
		\caption{Posterior distributions on the fraction of zero-spin \glspl{BBH}. We compare our \gls{TGMM}-based results with those of \cite{Tong2022}}
		\label{fig:TongComparisonFraction}
	\end{figure}
	
	As a demonstration of our method and its ability to reconstruct edge features with sufficient accuracy,
	we re-perform the analysis by \cite{Tong2022} \citep[see also][]{Galaudage2021,Callister:2022qwb} looking for a sub-population of \glspl{BBH} with negligible spins
	(i.e., a model with a delta function spike at $\chi = 0$) using events from the GWTC-3 catalog. 
	\cite{Tong2022} utilized specialized parameter estimation runs which provide an unbiased estimate of the distributions at $\chi = 0$.
	
	The fraction $\eta_0$ of \glspl{BBH} in the non-spinning subpopulation probes the scale of a narrow population feature at the edge of parameter space, and so it is especially sensitive to our ability to capture such features.
	We find agreement between the posterior on $\eta_0$ inferred by \cite{Tong2022} 
	and our results when applying \gls{TGMM} methods to the same model.
	However, we find agreement while using the same posterior samples as described in Section~\ref{sec:Data}, without requiring tailored parameter inference.
		
	\subsection{Model details}
	
	We replicate the \texttt{Extended} model implemented by \cite{Tong2022}, where the mass and redshift sectors are similar to \cite{LVKPop}. 
	The spin magnitude is modeled as a mixture of a delta function spike at $\chi_1 = 0, \chi_2 = 0$ and a component where both $\chi_1$ and $\chi_2$ are \gls{IID} 
	according to a more extended, truncated distribution.
	However, rather than utilizing the Beta distributions of the \texttt{Extended} model, we use truncated Gaussians, so in the spin section our population model is

	\begin{equation}
		\label{eq:TrucnatedNormalZeroSpinModel}
		\begin{aligned}
			&p\left(\chi_{1,2} \mid \mu_\chi, \sigma_\chi, \lambda_0\right) \\ 
			&= \left(1-\lambda_0\right) \phi_{[0,1]}\left(\chi_1 \mid \mu_\chi, \sigma_\chi\right) 
			\phi_{[0,1]}\left(\chi_2 \mid \mu_\chi, \sigma_\chi\right)\\
			&\ \ +\lambda_0 \delta\left(\chi_1\right) \delta\left(\chi_2\right)\,
		\end{aligned}
	\end{equation}
	where $\lambda_0$ is the fraction of \glspl{BBH} with zero spins.
	
	The spin orientation is  similar to the \texttt{Default} model in \cite{LVKPop}, but with 
	a minimum cosine tilt angle $z_{\min}$. 
	In particular, 
	\begin{equation}
		\label{eq:SpinOrientationTong}
		\begin{aligned}
			&p\left(z_{1,2} \mid \xi_{\textrm{spin}}, \sigma_{\textrm{spin}}, z_{\min }\right)
			= \\
			&\xi_{\textrm{spin}} \phi_{[-1,1]}\left(z_1 \mid \sigma_{\textrm{spin}}, z_{\min }\right) \phi_{[-1,1]}\left(z_2 \mid \sigma_{\textrm{spin}}, z_{\min }\right)\\
			&+(1-\xi_{\textrm{spin}})\left(\frac{\Theta\left(z_1-z_{\min }\right)}{1-z_{\min }}\right)\left(\frac{\Theta\left(z_2-z_{\min }\right)}{1-z_{\min }}\right)\,.
		\end{aligned}
	\end{equation}
	The analytic integrals needed with the Heaviside function in Eq.~\eqref{eq:SpinOrientationTong} are also implemented analytically, analogously to Eq~\eqref{eq:the-generic-integral}. 
	
		An important step in implementing the \cite{Tong2022} analysis is to make sure that our prior assumptions are as similar as possible. 
	Emulating the prior choices of \cite{Tong2022} is straightforward in all sectors except spin magnitude. 
	There, the prior on the Beta distribution parameters is chosen such that the prior on the true mean $\mathbb{E}(\chi)$  and variance of the spin distribution $\operatorname{Var}(\chi)$ are flat, together with cuts to ensure the Beta distributions are nonsingular.
	To reproduce this
	we use JAX to autodifferentiate expressions of $\mathbb{E}(\chi), \operatorname{Var}(\chi)$ and compute the relevant Jacobian factor
	to achieve flat priors on these. 
	We find that cuts on the Beta parameters are
	not needed in the case of truncated normal distributions.	
	Furthermore, as part of their analysis, \cite{Tong2022} ignored selection effects in the spin sectors. 
	As such, we modify our analysis by artificially inflating the relevant scale parameters in each \gls{TGMM} component in our fits to the selection injections.
	
	The result is that each component is essentially a uniform distribution in the spin sectors.

	Finally, there is a subtlety in the use of zero-spin parameter estimation samples in this analysis.
	In essence using such samples means that for the zero-spin subpopulation, the spin-orientations are not modeled (effectively creating a uniform distribution in $z_i = \cos \theta_i$). 
	In fact, as $\chi \to 0$, the orientation matters less and less and if we had infinite fidelity in our ability to reconstruct the posterior the likelihood distribution of the spin tilts would become uniform and there could be no information on the spin-orientation population in this limit.
	However, in practice we cannot use Eq.~\eqref{eq:SpinOrientationTong} for both subpopulations and expect to agree with the spin-orientation distributions used by \cite{Tong2022}.
	Instead, to make a fair comparison with our approach (and others that attempt to extrapolate to the zero-spin population, \citealt[c.f.][]{Callister:2022qwb})
	we must modify our spin-orientation model for the zero-spin subpopulation.
	As such, our complete spin model is
	\begin{align}
		p(\chi_1, \chi_2, z_1, z_2 \mid \Lambda) &=  \left(1-\eta_0\right) p(\chi_{1,2} \mid \mu_\chi, \sigma_\chi, \lambda_0 = 0)  \nonumber \\
		&\times p(z_{1,2} \mid \xi_{\textrm{spin}}, \sigma_{\textrm{spin}}, z_{\min }) \nonumber \\
		&+ \frac{1}{4} \eta_0 \delta(\chi) \,.
	\end{align}
	
	\subsection{Results}
	
	\begin{figure}[tb!]
		\begin{center}
			\includegraphics[width=0.4\paperwidth]{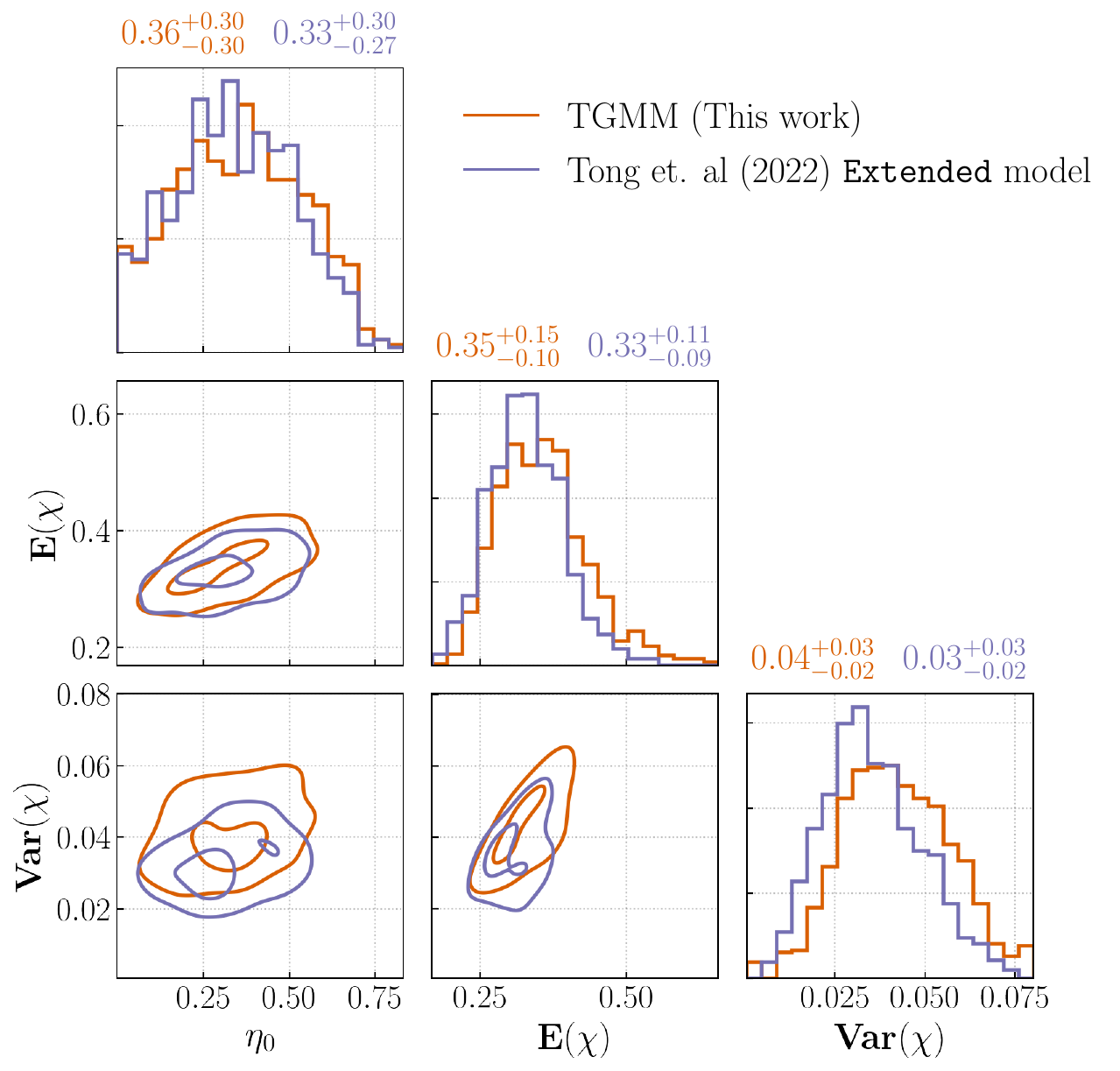}
		\end{center}
		\caption{Corner plot depicting the inferred fraction of zero-spin \glspl{BBH} as well as the mean and variance of the spinning subpopulation.
		Note that since the spin magnitude distributions have a different functional form, aside from $\eta_0$ we cannot directly compare the hyperparameters in this sector. 
		The isocontours representing the 50\% and 90\% credible intervals of the 2D marginals are given, along with the median and upper and lower bounds of the 90\% credible interval on the 1D marginals. 
		We show our results along with those of \cite{Tong2022}}
		\label{fig:TongComparisonSpinMagnitude}
	\end{figure}

\begin{figure}[tb!]
		\begin{center}
			\includegraphics[width=0.4\paperwidth]{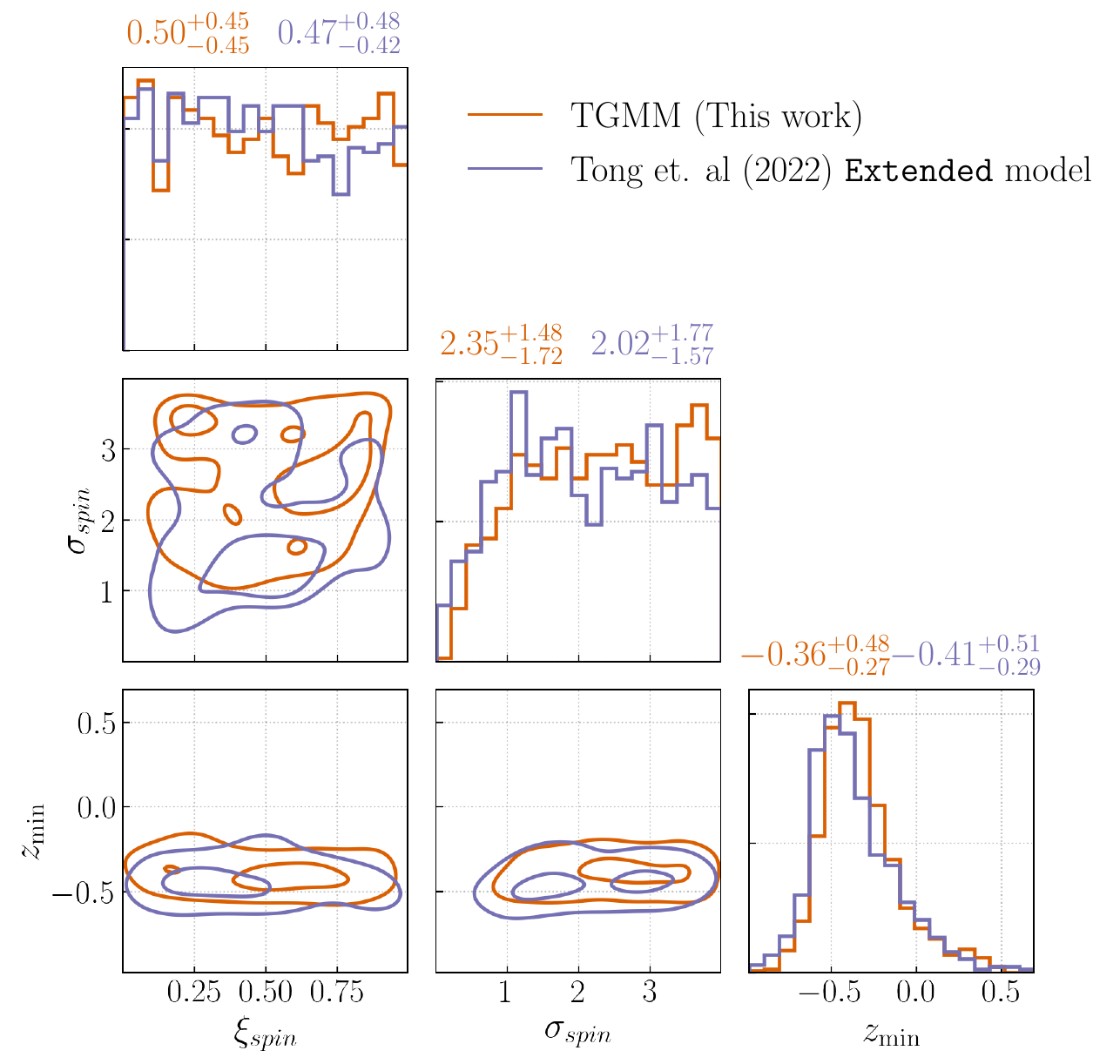}
		\end{center}
		\caption{As in Figure~\ref{fig:TongComparisonSpinMagnitude}, but illustrating the hyperparameters of the spin-orientation model.
		}
		\label{fig:TongComparisonSpinOrientation}
	\end{figure}

	Figure~\ref{fig:TongComparisonFraction} shows that our method reproduces the posterior of the fraction $\eta_0$ of zero-spin \glspl{BBH} quite well.
	In fact Figures~\ref{fig:TongComparisonSpinMagnitude} and ~\ref{fig:TongComparisonSpinOrientation} show a comparison of the corner plots in the spin magnitude and spin orientation population parameters. 
	These show that we reproduce the entire spin sector accurately, solely using parameter estimation samples where the \glspl{BH} are allowed to spin. 
	We also find agreement in the mass and redshift sectors of the population model, and give these comparisons in Appendix~\ref{sec:AdditionalResults}.
	
	\section{Conclusions}
	\label{sec:Conclusions}
	In this work, we introduced and validated a new approach for inferring populations of astrophysical objects whose parameters are bounded by physical constraints. Such boundary effects frequently arise, for example, in black hole spin magnitudes confined to $[0, 1]$ or exoplanet eccentricities within the same interval. Traditional estimators often struggle in reconstructing sharp features at the edges --- either suffering from high variance in the case of \gls{MC} methods or high biases through edge effects often introduced by density estimation techniques.
		
	We addressed these issues using a truncated Gaussian mixture model (TGMM) framework. By fitting samples to a mixture of truncated Gaussians, our method accurately recovers features near the boundary, while substantially mitigating variance blow-ups. We further showed how to combine TGMMs with standard hierarchical Bayesian population analyses: in the “analytic sector,” where the parameter is bounded and exhibits possible edge features, TGMM integrals can be computed analytically, whereas elsewhere we can fall back on more conventional \gls{MC} based methods.
	
	Through controlled 1D examples, we demonstrated that TGMMs sharply outperform naive Monte Carlo estimators near boundaries, maintaining high accuracy with comparatively fewer samples. Additionally, we incorporated our TGMM method into a realistic population inference analysis for binary black holes from the GWTC-3 catalog -- where black hole spins are a bounded domain of interest. We compared results of the analysis using both \gls{MC} techniques and our \gls{TGMM} method, and were able to get agreement between them especially in the spin posterior predictive distributions. 
	
	Our technique also probed near-zero spin BBH fractions from the same catalog -- a feature where naive \gls{MC} methods may fail -- without additional specialized parameter-estimation runs, validating that TGMM-based estimates track closely with results obtained using more computationally expensive methods.
	
	Not requiring additional parameter-estimation runs is a crucial advantage of our method.
	Firstly, this can be difficult to scale as the number of events grows in size.
	More importantly, in re-doing parameter estimation, the exact subpopulation distribution one would like to to probe needs to be decided apriori. One would like to be able to infer the sub-population's properties and its fraction simultaneously. Given the accuracy of our estimates for narrow features, we are able to do this and reproduce known results. 
	This allows us to let the data guide us towards the existence of certain subpopulations as opposed to modelling them explicitly. 	
	While TGMMs significantly reduce boundary-induced biases and variance issues, they do require that the population model separates into (i) a tractable part that can be expressed as a sum of truncated Gaussians, and (ii) a “sampled sector” for remaining parameters. This decomposition can still capture realistic physics in many scenarios, but certain highly correlated, multi-parameter distributions may require additional care. 
	 
	 For an end user, the procedure has been streamlined by a suite of packages, and the analysis moves forward as follows. To start, one identifies some subset or a sector of bounded parameters, and defines a population model in that sector of one of the general classes we mention (for example, the Truncated Gaussian), while defining something standard in the rest. Once done, we fit the posterior distributions and selection injections to \glspl{TGMM} (using \truncatedgaussianmixtures \citep{truncatedgaussianmixtures}) and utilize those data products as part of the inference procedure (performed using \gravpop \citep{gravpop}). 
	 One can then run the inference as normal, and the procedures described in the paper are handled by the package. In this sense \gravpop is one among many other packages that perform gravitational wave population inference (e.g. \textsc{GWPopulation} \citep{Talbot2025},  \textsc{GWInferno} \citep{Edelman_2023,git_gwinferno}, \textsc{ICAROGW} \citep{icarogw_paper, git_icarogw},  \textsc{PixelPop}\citep{Heinzel_2025} and \textsc{GWKokab} \citep{git_gwkokab, qazalbash2025gwkokabimplementationidentifyproperties} among others), but with the additional ability to probe narrow and/or edge-dominant population features.

	\begin{acknowledgements}
		We thank Will Farr for helpful discussions.
		AZ and AH were supported by NSF Grants PHY-2207594 and NSF Grant PHY-2308833 during the course of this work.
		The Flatiron Institute is a division of the Simons Foundation.
		This research has made use of data or software obtained from the Gravitational Wave Open Science Center (gwosc.org), a service of the LIGO Scientific Collaboration, the Virgo Collaboration, and KAGRA. This material is based upon work supported by NSF's LIGO Laboratory which is a major facility fully funded by the National Science Foundation, as well as the Science and Technology Facilities Council (STFC) of the United Kingdom, the Max-Planck-Society (MPS), and the State of Niedersachsen/Germany for support of the construction of Advanced LIGO and construction and operation of the GEO600 detector. Additional support for Advanced LIGO was provided by the Australian Research Council. Virgo is funded, through the European Gravitational Observatory (EGO), by the French Centre National de Recherche Scientifique (CNRS), the Italian Istituto Nazionale di Fisica Nucleare (INFN) and the Dutch Nikhef, with contributions by institutions from Belgium, Germany, Greece, Hungary, Ireland, Japan, Monaco, Poland, Portugal, Spain. KAGRA is supported by Ministry of Education, Culture, Sports, Science and Technology (MEXT), Japan Society for the Promotion of Science (JSPS) in Japan; National Research Foundation (NRF) and Ministry of Science and ICT (MSIT) in Korea; Academia Sinica (AS) and National Science and Technology Council (NSTC) in Taiwan.
	\end{acknowledgements}
	
	\appendix
	
	\section{Computation of $F$ with differing bounding limits}
	Here we provide the general analytic expression for the integral of a product of two truncated multivariate Gaussians, with possibly different boundary limits. Consider one truncated multivariate normal distribution with parameters $\bmu_1, \bSigma_1$, bounded within a hypercube defined by the lower corner $\ba_1$ and upper corner $\bb_1$, and another with parameters $\bmu_2, \bSigma_2$ and boundaries $\ba_2 \bb_2$. Then the integral of the product of the two pdfs is given by,
	\label{sec:different-bounded-corners}
		\begin{equation}
		\label{eq:AnalyticIntegralF}
		\begin{aligned}
			&F_{[\ba_1, \bb_1; \ba_2, \bb_2]}(\bmu_1, \bSigma_1, \bmu_2, \bSigma_2)  =\\
			& \int_{\ba}^{\bb} \phi_{[\ba_1, \bb_1]}(\btheta | \bmu_1 ,\bSigma_1) 
			\phi_{[\ba_2, \bb_2]}(\btheta \mid \bmu_2 ,\bSigma_2)d\btheta \\
			&= \frac{C_{[\ba_3, \bb_3]}(\bmu_{3}, \bSigma_{3})}
			{C_{[\ba_1, \bb_1]}(\bmu_1, \bSigma_1)C_{[\ba_2, \bb_2]}(\bmu_2, \bSigma_2)} 
			\phi(\bmu_1 \mid \bmu_2,\ \bSigma_1 + \bSigma_2)\,.
		\end{aligned}
	\end{equation}
	Here
	\begin{equation}
		\begin{aligned}
			C_{[\ba,\bb]}&(\bmu, \bSigma) = \int_{\ba}^{\bb} \phi(\btheta \mid \bmu, \bSigma) d\btheta \,, \\
			\bmu_3& =\left(\boldsymbol{\Sigma}_1^{-1}+\boldsymbol{\Sigma}_2^{-1}\right)^{-1}
			\left(\boldsymbol{\Sigma}_1^{-1} \bmu_1+\boldsymbol{\Sigma}_2^{-1} \bmu_2\right) \\
			\bSigma_3& =\left(\boldsymbol{\Sigma}_1^{-1}+\boldsymbol{\Sigma}_2^{-1}\right)^{-1}\, \\
			\ba_3& =\max(\ba_1, \ba_2) \\
			\bb_3&=\min(\bb_1, \bb_2)\,,
		\end{aligned}
	\end{equation}
	where $\phi$ is a Gaussian distribution (without truncation) and the $\max(\cdot)$ and $\min(\cdot)$ are taken element wise. 
	We describe the numerical implementation details of $C_{\ba,\bb}(\bmu, \bSigma)$ in Appendix~\ref{sec:ComputationOfIntegral}.

	\section{Variance of Monte-Carlo Estimation for 1D Example}
	\label{sec:MCVariance}

	The variance of the \gls{MC} estimator in Section~\ref{sec:numerical-comparison-of-methods} can be calculated analytically for our choice of population model, and the shape of the posterior. 
	In general we have that
	\begin{equation}
		\label{eq:variance_mc_estimate}
	\begin{aligned}
		\operatorname{Var}[\mathcal{\hat{L}}^{MC}_N] &= \frac{1}{N} \left[\langle p(\theta \mid \Lambda)^2 \rangle_{\theta \sim p(\theta)} - \langle p(\theta \mid \Lambda) \rangle^2_{\theta \sim p(\theta)}\right]\\
		&= \frac{1}{N} \left[\langle p(\theta \mid \Lambda)^2 \rangle_{\theta \sim p(\theta)} - \mathcal{L}^2\right]\,,
	\end{aligned}
	\end{equation}
	where $p(\theta \mid \Lambda)$ is our population model and $p(\theta)$ is our posterior. 
	We calculate both terms in Eq.~\eqref{eq:variance_mc_estimate},analytically below. 
	In our toy example our population model is
	\begin{equation}
	p(\chi \mid \mu, \sigma) = \phi_{[0,1]}(\chi \mid \mu, \sigma)\,,
	\end{equation}
	and our posterior is also a truncated normal which for now we represent as
	\begin{equation}
	p(\chi) = \phi_{[0,1]}(\chi \mid \mu_{0}, \sigma_{0})\,,
	\end{equation}
	generalizing beyond $\mu_0 = 0$.
	The expected value of the \gls{MC} estimate is the basis of the second term in Eq.~\eqref{eq:variance_mc_estimate}, and can be computed analytically as,
	\begin{equation}
		\begin{aligned}
		\langle p(\theta \mid \Lambda) \rangle_{\theta \sim p(\theta)}
		&= \int_0^1 \phi_{[0,1]}(\theta \mid \mu, \sigma)\phi_{[0,1]}(\theta \mid \mu_0, \sigma_0) d\theta \\
		&=C_{[0,1]}\left(\frac{\mu\sigma_0^2 + \mu_0 \sigma^2}{\sigma_0^2 + \sigma^2}, \frac{\sigma_0\sigma}{\sqrt{\sigma_0^2 + \sigma^2}}\right) \phi(\mu_0 \mid \mu, \sqrt{\sigma_0^2 + \sigma^2})\,,
		\end{aligned}
	\end{equation}
	where $C_{[0,1]}(\mu, \sigma)$ denotes the probability mass of a Gaussian of mean $\mu$ and standard deviation $\sigma$, truncated over $[0,1]$.
	We can also get the first term in Eq.~\eqref{eq:variance_mc_estimate} as the second moment of the population model under the posterior distribtion, and the analytic form of the term is
	\begin{equation}
	\begin{aligned}
	\langle p(\theta \mid \Lambda)^2 \rangle_{\theta \sim p(\theta)} 
	&=\int_0^1 \phi_{[0,1]}(\theta \mid \mu, \sigma)^2 \phi_{[0,1]}(\theta \mid \mu_0, \sigma_0) d\theta 
	=\int_0^1 \frac{\phi(\theta \mid \mu, \sigma)^2 \phi(\theta \mid \mu_0, \sigma_0)}{C_{[0,1]}(\mu, \sigma)^2C_{[0,1]}(\mu_{0}, \sigma_{0})} d\theta \\
	&=\frac{C_{[0,1]}\left(
		\frac{2\mu \sigma_0^2 +\mu_0\sigma^2}{2\sigma_0^2 + \sigma^2}, 
		\frac{\sigma_0 \sigma}{\sqrt{2\sigma_0^2 + \sigma^2}}
		\right)}
		{C_{[0,1]}(\mu, \sigma)^2 
		C_{[0,1]}(\mu_{0}, \sigma_{0})}
		\frac{\phi\left(\mu_0 \mid \mu, \sqrt{\sigma_0^2 + \frac{\sigma^2}{2}}\right)}
		{\sqrt{2\pi}\sigma}\,.
	\end{aligned}
	\end{equation}
	This unweildy expression is used to calculate the 95\% confidence intervals in Figure~\ref{fig:IllustrativeExample}.

	\section{\gls{KDE} with $\mathcal{O}(\lowercase{h})$ boundary bias}
	\label{sec:BoundaryKDE}

	In this section we derive a bounded \gls{KDE} method which is unbiased to $\mathcal{O}(h)$ at the boundary.
	The \gls{KDE} $\hat{p}(x)$ of a particular probability distribution is given by
	\begin{equation}
		\begin{aligned}
			\hat{p}(x) = \frac{1}{N}\sum_i K(x | x_i , h)\,,
		\end{aligned}
	\end{equation}
	where $K(x | x_i, h)$ is a (generally non-symmetric) kernel with a bandwidth $h$. 
	An example would be a Gaussian KDE, in which case $K(x | x_n, h) = \frac{1}{h}\phi((x-x_n)/h)$. 
	However, we are particularly interested in representing distributions on truncated domains $x \in [a,b]$.
	In the large $N$ limit this estimate becomes
	\begin{equation}
		\begin{aligned}
			\hat{p}(x) = \mathbb{E}_{x' \sim p(x)}[K(x|x')]
		\end{aligned}
	\end{equation}
	We would like that in this limit the estimate becomes the true distribution, up to terms of order $\mathcal{O}(h)$:	
	\begin{equation}
		\begin{aligned}
			\mathbb{E}[\hat{p}(x)] = \int K(x | x')p(x')dx'\,.
		\end{aligned}
	\end{equation}
	
	Assume we can analytically get $p(x')$ by Taylor expanding around $x$. 
	This should be possible since $x,x' \in [a,b]$ and it is a fair assumption that the PDF we want to fit to is analytic inside the truncated domain. 
	Then
	\begin{equation}
		\begin{aligned}
			\mathbb{E}[\hat{p}(x)] 
			& = \int K(x | x')
			\left[ p(x) + h\left(\frac{x-x'}{h}\right)
			\cdot\grad p(x) + \mathcal{O}(h^2)   \right]dx' 
			\\
			& = p(x)\int K(x | x')dx' + \grad p(x)\cdot \vec{x}\int K(x | x')dx' - \grad p(x)\cdot\int \vec{x}'K(x | x')dx' + \mathcal{O}(h^2)\,.
		\end{aligned}
	\end{equation}

	This shows that an unbiased estimate at $x$, of order $\mathcal{O}(h)$ requires
	\begin{equation}
		\begin{aligned}
			\int K(x | x')dx' & = 1,  
			& \text{and} && \int \vec{x}'K(x | x')dx' &= \vec{x} + \mathcal{O}(h) \,.
		\end{aligned}
	\end{equation}
	We would further like our kernels to be of the form of a truncated normal. 
	More specifically, 
	\begin{equation}
		\begin{aligned}
			K(x | x' h) = \phi_{[a,b]}(x\mid x', h) = \frac{\phi(x | x', h)}{C(x', h)} \,.
		\end{aligned}
	\end{equation}
	However this kernel is definitely not unbiased as we get close to the edge, since
	\begin{equation}
		\begin{aligned}
			\int K(x | x')dx' = \int \frac{\phi(x | x' h)}{C(x', h)}dx' \neq 1 \,.
		\end{aligned}
	\end{equation}
	
	However, there is a way to remedy this while retaining truncated Gaussians for our kernels. 
	If we take the mean parameter of the kernel $\mu$ , and instead of placing it at the point $x'$ it to be some function of the point $x'$, we may have enough degrees of freedom to make it an unbiased KDE. 
	This makes our ansatz for the kernel $K(x | x') = \phi_{[a,b]}(x | \mu(x') h)$. 
	The goal is then to find the function $\mu(x')$ to give us an unbiased estimate of $\hat p(x)$.
		
	Using our ansatz, let us assume we can find a function $x'(\mu)$ such that $dx'/d\mu = C(\mu, h)$, which then gives
	\begin{equation}
		\begin{aligned}
			\int K(x | x')dx' = \int_{-\infty}^{\infty} \frac{\phi(x | \mu, h)}{C(\mu, h)} \frac{dx'}{d\mu} d\mu = \int_{-\infty}^{\infty} \phi(x | \mu, h) d\mu = 1 \,.
		\end{aligned}
	\end{equation}
	This means that in the 1D case that
	\begin{equation}
		\begin{aligned}
			x'(\mu) = \int C(\mu, h) d\mu 
			= \int \Phi\left(\frac{b-\mu}{h}\right) 
			- \Phi\left(\frac{a-\mu}{h}\right) d\mu \,.
		\end{aligned}
	\end{equation}
	where $\Phi$ is the CDF of a gaussian distribution.
	This integral of a gaussian CDF is well known and can be computed to give
	\begin{equation}
		\begin{aligned}
			x'(\mu) = b - h \left( \beta \Phi(\beta) + \phi(\beta) \right) + h \left[ \alpha \Phi(\alpha) + \phi(\alpha) \right] \,,
		\end{aligned}
	\end{equation}
	where $\alpha = (a-\mu)/h$ and $\beta = (b-\mu)/h$.
	One can invert this equation for any given $x'$ to find the $\mu$ parameter for the truncated Gaussian kernel.
	Inversion using root finding is very cheap in this situation and even for $100,000$ points, it is not a limiting factor in the speed of the \gls{TGMM} fitting procedure described in Section~\ref{sec:GeneralProcedure}. 
	Interpolation is also a good way to cache results and scale this procedure but we did not find that this step needed any further optimization for our use case. 
	
	As for the second condition for the kernel to be unbiased, we evaluate the moment
	\begin{equation}
		\begin{aligned}
			\int x'K(x | x')dx' & = \int_a^b x'\frac{\phi(x | \mu(x') h)}{C(\mu(x'), h)}dx' 
			= \int_{-\infty}^{\infty} x'(\mu)\frac{\phi(x | \mu, h)}{C(\mu, h)} \frac{dx'}{d\mu} d\mu 
			= \frac{1}{\sqrt{2\pi}h}\int_{-\infty}^{\infty} x'(\mu)\exp(-\frac{(x-\mu)^2}{2h^2})d\mu
			\\
			& = \mathbb{E}_{\mu \sim \mathcal{N}(x,h)}\left[x'(\mu) \right] 
			= \mathbb{E}_{\mu \sim \mathcal{N}(x,h)}\left[\mu + (x'(\mu) - \mu) \right] 
			=  \mathbb{E}_{\mu \sim \mathcal{N}(x,h)}\left[\mu \right] + \mathbb{E}_{\mu \sim \mathcal{N}(x,h)}\left[ (x'(\mu) - \mu) \right]\,.
		\end{aligned}
	\end{equation}
	
	\begin{figure}
		\label{fig:UnbiasedKDEScaling}
		\begin{center}
			\includegraphics[width=0.5\paperwidth]{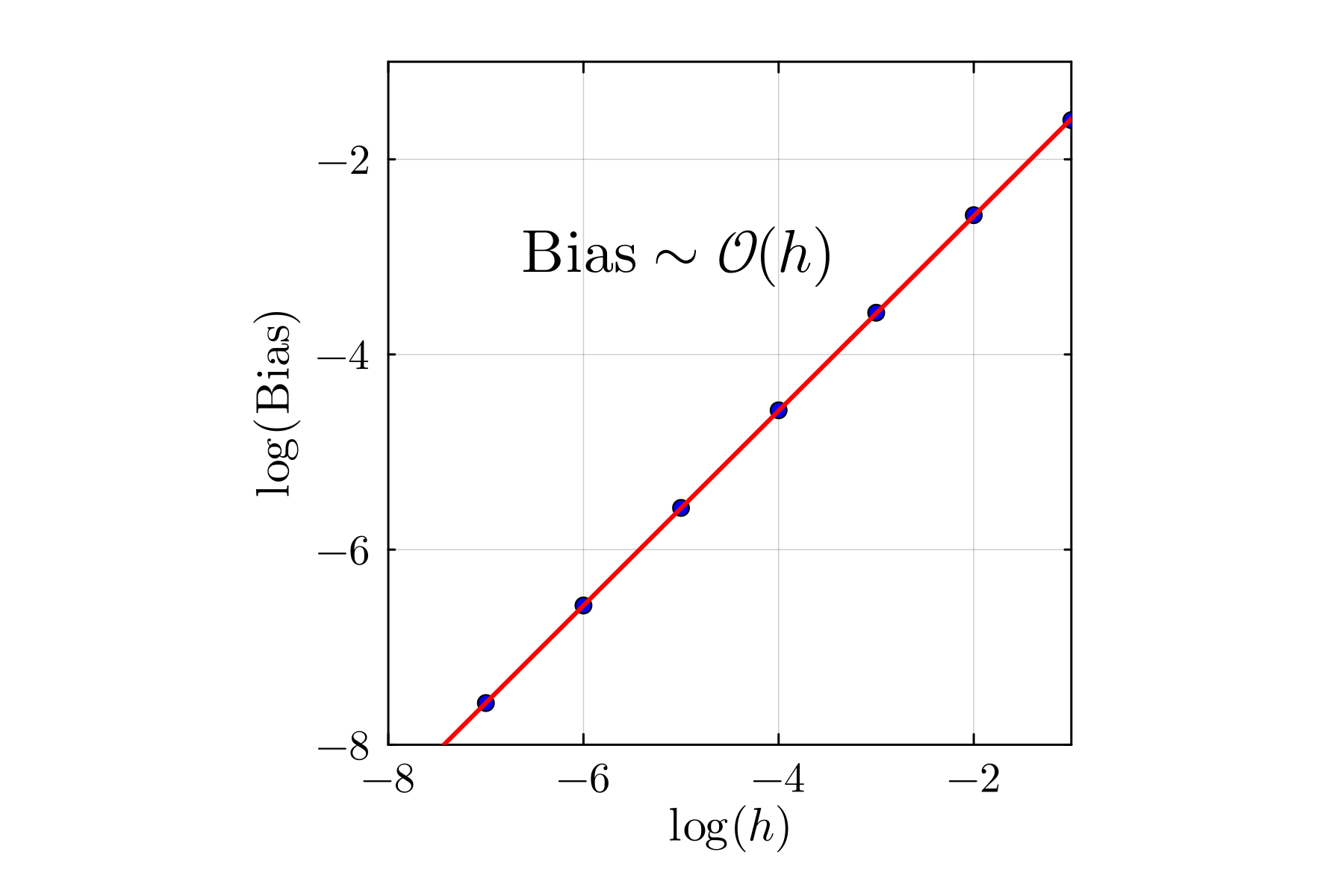}
		\end{center}
		\caption{We plot the logarithm of the bias term in Eq.~\eqref{eq:KDEBoundarybias} against the logarithm of the bandwidth $h$. 
		A positive slope of $1$ is observed in the plot, which implies that the bias is of order $\mathcal{O}(h)$.}
	\end{figure}
	
To be able to perform this integral and see the scaling with $h$, we focus on the region near the left edge at $a$ let the right most limit $b \to \infty$.
This is because the bias should only occur close to the edges, and if $h$ is small the effect of the right edge on the bias at the left edge can be safely ignored. 

Hence we have
\begin{equation}
	\begin{aligned}
		\label{eq:KDEBoundarybias}
		\int x'K(x | x')dx' 
		&= x + \mathbb{E}_{\mu \sim \mathcal{N}(x,h)}\left[ (x'(\mu) - \mu) \right] 
		= x + \mathbb{E}_{\mu \sim \mathcal{N}(x,h)}\left[h \left(\phi(\alpha(\mu))+\alpha(\mu)  \Phi (\alpha(\mu) )\right)  \right] \\
		&=  x + \int_{-\infty}^{+\infty} \left[\phi(\alpha(\mu))+\alpha(\mu)  \Phi (\alpha(\mu))  \right] \phi\left(\frac{x-\mu}{h}\right) d\mu\, \\
		&= x + h \, I\left(\frac{x-a}{h}\right)\,,
	\end{aligned}
\end{equation}
where
\begin{equation}
	\begin{aligned}
		I(y) &= -\int_{-\infty}^{+\infty} \left[\phi(z)+z\Phi(z)  \right] \phi(y + z)dz \,.
	\end{aligned}
\end{equation}

Note that $(x-a)/h$ simply counts how many bandwidths away from the boundary we are. 
Close to the boundary ($x \to a$) we expect $I([x-a]/h)$ to be $\mathcal{O}(1)$ and hence  
\begin{equation}
	\int x'K(x | x')dx' = x + \mathcal{O}(h)
\end{equation}
as required. 
This proves that the KDE is unbiased at order $\mathcal{O}(h)$
We can numerically verify this scaling with $h$.
The results are shown on Fig.~\ref{fig:UnbiasedKDEScaling}. 
One can clearly see a slope of $\approx 1$, which emperically confirms the \gls{KDE} is indeed biased only at order $\mathcal{O}(h)$.
	
	\section{Truncated Gaussian Mixture Models}
	\label{sec:TGMM}
	Our estimate of the integral, \eqref{eq:hybrid-scheme}, relies on fitting the 
	weighted samples from each event posterior $p(\btheta | d_e)$ to a 
	\gls{TGMM}. 
	The estimate is only as good as the fit of the \gls{TGMM} to the data. 
	In this section we discuss the details of the methods and techniques we use for this fit.	
	The general algorithm to fit mixtures of multivariate truncated Gaussians to data is outlined in \cite{Lee2012} which we follow for our implementation in \cite{truncatedgaussianmixtures}. 
	We have added two features not implemented in \cite{Lee2012}. 
	Firstly, we add the ability to fit these \glspl{TGMM} to weighted samples. 
	The algorithm to fit these is outlined for the case of standard \glspl{GMM} is outlined in \cite{frisch2021gaussianmixtureestimationweighted}, 
	which we modify for our application. 
	Secondly \glspl{TGMM} can have slow convergence when fitting features near the boundary. 
	To side step this, we perform some initial iterations where we fit the \glspl{TGMM} directly to boundary unbiased \glspl{KDE} of the data, as opposed to samples.
	This speeds up the iterations taken for the \gls{TGMM} to converge to an edge feature. 
	We describe this modification to the algorithm here, but describe our boundary unbiased \gls{KDE} (which to our knowledge is novel) in Appendix~\ref{sec:BoundaryKDE}.
	
	We have released a \texttt{julia} package \href{https://github.com/Potatoasad/TruncatedGaussianMixtures.jl}{\texttt{TruncatedGaussianMixtures.jl}} \citep{truncatedgaussianmixtures} that can fit these mixtures to any user-provided dataset. 
	In addition we have released a wrapper package in \texttt{python} as \href{https://github.com/Potatoasad/truncatedgaussianmixtures}{\texttt{truncatedgaussianmixtures}}.
	
	\subsection{Full Algorithm}
	\
	We start with a set of weighted datapoints  $\mathcal{D} = \{\btheta_i, W_i\}_{i=1\dots N}$, with $W_i$ being the weight of the $i$th sample. 
	We follow the convention that $\sum_i W_i = 1$. 
	We aim to fit these samples to a mixture of $K$ truncated multivariate Gaussians with parameters $\bmu_k, \bSigma_k$, with weights $w_k$ truncated over a hypercube with corners ${\bf a}$ and ${\bf b}$. 
	For brevity we represent the set of parameters describing this mixture as $\bTheta = \{\bmu_k, \bSigma_k, w_k\}$.
	The outcome of our fit would be to infer $\bTheta$, which gives the best fit of the distribution that produced $\mathcal D$ as
	\begin{equation}
		p(\btheta\mid \mathcal \bTheta) = \sum_k^K w_k \phi_{[\ba, \bb]}(\btheta \mid \bmu_k, \bSigma_k)\,.
	\end{equation}
	
	The implementation of the expectation maximization
	algorithm introduces a set of unknown, latent, indicator variables ${\bf z}_{i} \in \{0,1\}^K$ representing the assignment of datapoints $\btheta_i$ to particular TGMM components $k$.
	\begin{align}
	\label{eq:CategoricalZ}
		[\mathbf{z}_i]_k =  \begin{cases}
			1 & \parbox[t]{0.4\linewidth}{if the $i$th datapoint is assigned to component $k$}, \\
			0 & \parbox[t]{0.4\linewidth}{if the $i$th datapoint is not assigned to component $k$}.
		\end{cases} \sim \text{Categorical}(\langle z_{ik}\rangle)\,,
	\end{align}
	where the distribution is completely characterized by the responsibility matrix $\langle z_{ik}\rangle$.
	The responsibility matrix can be understood by noting that for each data point $i$, there is a $k$ dimensional vector  $[\boldsymbol{\eta}_i]_k = \langle z_{ik}\rangle$ representing the probability that this point belongs to the $k$th component, 
	\begin{equation}
		p({\bf z}_{i} = \mathbbm{1}_k) = \langle z_{ik}\rangle\,,
	\end{equation}
	since the expectation of an indicator variable in general is its own probability mass function.
	
	Having knowlege of the latent variables allows us to make hard assignments to each component, and write out our likelihood as\footnote{
	The addition of the $W_i$ to Eq.~\eqref{eq:full-EM-likelihood} is consistent since for a datapoint with zero weight assigned to it, we would like our model to attempt to return a zero probability in that location regardless of the component it is assigned to.}
	\begin{equation}
		\label{eq:full-EM-likelihood}
		\begin{aligned}
			p(\btheta_i  \mid {\bf z}_i = \mathbbm{1}_k ,\bTheta) &= \phi_{[\ba, \bb]}(\btheta_i \mid \bmu_k, \bSigma_k)\,.
		\end{aligned}
	\end{equation}
	Eq.~\eqref{eq:full-EM-likelihood} simply points out that the point $i$ follows the distribution of component $k$ conditioned on the fact that it is assigned to component $k$.
	Using bayes rule we get,
	\begin{equation}
		\begin{aligned}
			p(\btheta_i , {\bf z}_i = \mathbbm{1}_k \mid \mathcal  \bTheta) &= p(\btheta_i  \mid {\bf z}_i = \mathbbm{1}_k, \mathcal  \bTheta) p({\bf z}_i = \mathbbm{1}_k \mid \bTheta)\\
			&= \phi_{[\ba, \bb]}(\btheta_i \mid \bmu_k, \bSigma_k)\langle z_{ik} \rangle\,,
		\end{aligned}
	\end{equation}
	where we define 
	\begin{equation}
		\langle z_{ik} \rangle = p({\bf z}_i = \mathbbm{1}_k \mid \bTheta) = \mathbb{E}(z_{ik} \mid \bTheta)
	\end{equation}
	
	The likelihood we seek to optimize is the expected value of the log-likelihood over the weighted dataset,
	\begin{equation}
		\log\mathcal{L}(\bTheta \mid \mathcal D) =  \sum_i^N W_i \log \left[  \sum_k^K w_k \phi_{[\ba, \bb]}(\btheta_i \mid \bmu_k, \bSigma_k) \right] \,.
	\end{equation}
	Of course, we do not yet know $w_k$, since it is what we are trying to infer. 
	To get around this we start with some initial set of TGMM parameters $\bTheta^{(0)}$, compute $\langle z^{(0)}_{ik} \rangle = \mathbb{E}(z_{ik} \mid \bTheta^{(0)})$, and insert them into the expression,
	\begin{equation}
		\begin{aligned}
			&\log\mathcal{L}(\bTheta \mid \mathcal D, \bTheta^{(0)}) = \\ &
			\sum_i^N W_i \log\left[\sum_k^K \langle z^{(0)}_{ik} \rangle  \frac{w_k \phi_{[\ba, \bb]}(\btheta_i \mid \bmu_k, \bSigma_k)}{\langle z^{(0)}_{ ik} \rangle} \right]\,.
		\end{aligned}
	\end{equation}
	The initialziation can be performed using K-means clustering, as is customary with initialization methods for \glspl{GMM} \citep{Lloyd:1982zni, 2020SciPy-NMeth}.
	
	The term inside the logarithm is an expectation over ${\bf z}_i \mid \bTheta^{(0)}$. We then use Jensen's inequality to exchange the logarithm and the expectation to get a lower bound on the log-likelihood of our estimate, which we denote as $Q(\bTheta \mid \bTheta^{(0)}, \mathcal D) $,
	\begin{equation}
		\label{eq:introduceQ}
		\begin{aligned}
			&\log\mathcal{L}(\bTheta \mid \mathcal D, \bTheta^{(0)}) \geq
			Q(\bTheta \mid \bTheta^{(0)}, \mathcal D) =  \sum_i^N \sum_k^K W_i \langle z^{(0)}_{ ik} \rangle \log \left[\frac{ w_k \phi_{[\ba, \bb]}(\btheta_i \mid \bmu_k, \bSigma_k)}{\langle z^{(0)}_{ ik} \rangle} \right]\,.
		\end{aligned}
	\end{equation}
	To make the inequality in Eq.~\eqref{eq:introduceQ} an equality, we want the term inside the logarithm to be as close to a constant as possible, and so we make the closest guess we can with the initial set of parameters,
	\begin{equation}
		\label{eq:Estep-precursor}
		\langle z_{ik}^{(0)}\rangle \propto  w^{(0)}_k \phi_{[\ba, \bb]}(\btheta_i \mid \bmu^{(0)}_k, \bSigma^{(0)}_k)\,.
	\end{equation}
	Since $\sum_k \langle z_{ik}^{(0)}\rangle = 1$, we can identify the normalization constant and state the first step in the EM algorithm, called the E-step.
	
	\textbf{The E-Step: } $\langle z_{ik}^{(0)} \rangle$ can be computed from the initial parameter guess $\bTheta^{(0)}$ as
	\begin{equation}
		\label{eq:Estep}
		\langle z_{ik}^{(0)}\rangle = \frac{w^{(0)}_k \phi_{[\ba, \bb]}(\btheta_i \mid \bmu^{(0)}_k, \bSigma^{(0)}_k)}{\sum_k w^{(0)}_k \phi_{[\ba, \bb]}(\btheta_i \mid \bmu^{(0)}_k, \bSigma^{(0)}_k)}\,,
	\end{equation}
	which is a normalized version of Eq.~\eqref{eq:Estep-precursor}.
	
	\textbf{The M-Step: }We then try and maximize $Q(\bTheta \mid \bTheta^{(0)})$ to get a new estimate for the maximum $\bTheta^{(1)}$,

	\begin{equation}
		\begin{aligned}
			& \bTheta^{(1)} = \arg \max_{\bTheta} Q(\bTheta \mid \bTheta^{(0)}, \mathcal D) 
			\\ 
			&= \arg \max_{\bTheta} \left[\sum_i^N \sum_k^K W_i \langle z_{ik}^{(0)} \rangle \log \left[w_k \phi_{[\ba, \bb]}(\btheta_i \mid \bmu_k, \bSigma_k)\right]\right],
		\end{aligned}
	\end{equation}
	subject to the condition that $\sum_k w_k = 1$, and ignoring constant terms independent of $\bTheta$. 
	
  	Using this constraint and the separation of the logarithms, we can first find the new estimate of $w_k$ as	
  	\begin{equation}
		w^{(1)}_k = \frac{\sum_{i}W_i \langle z_{ik}^{(0)} \rangle }{\sum_{i}W_i}\,.
	\end{equation}
	With a fixed estimate for $w_k$ in the M-step, we can write the inference of the remaining parameters ($\bmu_k$ and $\bSigma_k$) as $K$ separate maximization procedures, and find that
	\begin{equation}
		\begin{aligned}
			\bTheta^{(1)}_k = \arg \max_{\bTheta_k}  \left[\sum_i^N \eta_{ik} \log \left[\phi_{[\ba, \bb]}(\btheta_i \mid \bmu_k, \bSigma_k)\right]\right]\,,
		\end{aligned}
	\end{equation}
	where we have removed constant terms\footnote{This includes $w_k$ since these are now fixed.} and defined weights for each sample as
	\begin{equation}
		\eta_{ik} := \frac{W_i \langle z_{ik}^{(0)} \rangle}{\sum_{i} W_i \langle z_{ik}^{(0)} \rangle}\,.
	\end{equation}
	Let $\mathcal{D}_k$ be the weighted dataset $\{\btheta_i,  \eta_{ik}\}_{i=1\dots N}$. 
	Then the sum can be written as the expectation
		\begin{align}
			\bTheta^{(1)}_k &= 
			\arg \max_{\bTheta_k}  \left[\mathbb{E}_{\btheta \sim \mathcal{D}_k} 
			\left[\log\phi_{[\ba, \bb]}(\btheta \mid \bmu_k, \bSigma_k)\right]\right] \\
			&\overset{\mathrm{c}}{=}  
			\arg \min_{\bTheta_k} D_{KL}(\mathcal{D}_k\ ||\  \mathcal{N}_{[\ba,\bb]}(\bmu_k, \bSigma_k))\,,
		\end{align}
	where $\overset{\mathrm{c}}{=}$ denotes equality up to an additive constant. This expectation is simply the KL-divergence between $\mathcal{D}_k$ and $\mathcal{N}_{[\ba,\bb]}(\bmu_k, \bSigma_k)$, up to terms independent of $\bmu_k$ and $\bSigma_k$.
	For forms of the target distribution that are in the exponentional family (of which truncated Gaussians are a member), minimizing the KL divergence corresponds to matching the sufficient statistics of the two distributions \citep{amari2000methods}. 
	Hence we simply chose to find $\bmu_k$ and $\bSigma_k$ that solve the simultaneous equations involving the sufficient statistics of the truncated Gaussian family---namely the first two moments:
		\begin{align}
			\mathbb{E}_{\btheta \sim \mathcal{N}_{[\ba,\bb]}(\bmu_k, \bSigma_k)}[\btheta] &=\sum_i \eta_{ik} \btheta_i \,, \\
			\mathbb{E}_{\btheta \sim \mathcal{N}_{[\ba,\bb]}(\bmu_k, \bSigma_k)}[\btheta\btheta^T] &= \sum_i \eta_{ik} \btheta_i\btheta_i^T\,.
			\label{eq:moment-matching}
		\end{align}
	The algorithm to compute the left hand side of Eq.~\eqref{eq:moment-matching} is outlined in \cite{Lee2012} and implemented in our package \cite{truncatedgaussianmixtures}. 
	This matching can then be done by using a fixed point iteration scheme using auto-differentiation techniques to get the required gradients (though some custom gradient rules are required). 
	This iteration scheme is also implemented in~\cite{truncatedgaussianmixtures}.
		
		\begin{algorithm}[H]
			\caption{{\it Expectation Maximization Routine for \glspl{TGMM} with customizations.}
				We take in our weighted dataset of samples and weights $\mathcal{D} = \{\btheta_i, W_i\}_{i=1\dots N}$, 
			 and aim to fit to this data $K$ gaussians truncated to lie in a hypercube with corners $[\ba, \bb]$. 
				Additionally we initialize a projection operator $\mathcal{P}$, that projects a covariance matrix to some user-specified block structure. 
				We define a tolerance \texttt{tol} to provide a stopping criteria for the expectation maximization.}
			\label{alg:EM}
			\begin{algorithmic}[1]
				\State{Initialize: \texttt{K-means}($\btheta_i$, $W_i$) $\to z^{(0)}_{ik} \in \{0,1\}$.}
				\State{Use to define initial parameters as 
					\begin{equation}
						\begin{aligned}
							&\eta^{(0)}_{ik} = \frac{W_i z^{(0)}_{ik}}{\sum_{i} W_i z^{(0)}_{ik}},\ \ \ 
							w^{(0)}_k = \frac{\sum_i W_i z^{(0)}_{ik}}{\sum_{i} W_i},\ \ \ 
							\bmu^{(0)}_k = \sum_{i} \eta^{(0)}_{ik} \btheta_i,\ \ \ 
							\bSigma^{(0)}_k = \sum_{i} \eta^{(0)}_{ik} \left(\btheta_i - \bmu^{(0)}_k\right)\left(\btheta_i - \bmu^{(0)}_k\right)^T 
							\\
							&\bTheta^{(0)} = \{w^{(0)}_k, \bmu^{(0)}_k, \bSigma^{(0)}_k\}
						\end{aligned}
				\end{equation}}
				\State{\textbf{Boundary unbiasing post-initialization steps: }}
				\State{Precompute first and second moments for the \gls{KDE} kernels, $\mathcal{M}_{1 i}$ and $\mathcal{M}_{2 i}$ (see Eq.~\eqref{eq:moment-matching-updated-kde}) for each point $\btheta_i$}
				\For{$n \in 1\dots N_{\text{KDE}}$}
				\For{$k \in 1\dots K$}
				\State{$\texttt{E-Step}(\bTheta^{(n-1)}) \to \langle z_{ik}^{(n)} \rangle$}
				\State{$\texttt{M-Step-KDE-variant}(\langle z_{ik}^{(n-1)} \rangle, \bTheta^{(n-1)}) \to \bTheta^{(n)}$} 
				\State{Project onto block structure $\bSigma^{(n)}_k \to \mathcal{P}\bSigma^{(n)}_k$} 
				\EndFor
				\EndFor
				\State{\textbf{Fitting: }}
				\State{Now that we have fit to the \gls{KDE} and allowed the PDF to fit edge features, we will allow it to converge with the standard M-steps}
				\For{$n \in N_{\text{KDE}}\dots N_{\text{iterations}}$}
				\For{$k \in 1\dots K$}
				\State{$\texttt{E-Step}(\bTheta^{(n-1)}) \to \langle z_{ik}^{(n)} \rangle$}
				\State{$\texttt{M-Step}(\langle z_{ik}^{(n-1)} \rangle, \bTheta^{(n-1)}) \to \bTheta^{(n)}$} 
				\State{Project onto block structure $\bSigma^{(n)}_k \to \mathcal{P}\bSigma^{(n)}_k$} 
				\If{$\Delta\ln\mathcal{L}(\bTheta^{(n)} \mid \mathcal D) \leq \texttt{tol}$}
				\State{\Return{$w^{(n)}_k, \bmu^{(n)}_k, \bSigma^{(n)}_k$}}
				\EndIf
				\EndFor
				\EndFor
				\State{\Return{$w^{(N_{\text{iter}})}_k, \bmu^{(N_{\text{iter}})}_k, \bSigma^{(N_{\text{iter}})}_k$}}
			\end{algorithmic}
		\end{algorithm}
	
	\subsection{Boundary Unbiasing}
	\label{sec:BoundaryUnbiasing}
	We find that when fitting \glspl{TGMM} to samples, getting better convergence at the edge is slow. 
	Since we we use a K-means initialization, most initialized components lie mostly within the truncation region, and hence the initialization itself is biased to begin with. 
	To counteract this, we look more closely at Eqs.~\eqref{eq:moment-matching} as part of the M-step in the algorithm.
	The right hand side of these equations are the weighted moments of the samples, which can be viewed as weighted moments of a PDF given by $\hat{p}_i(\btheta) = \sum_i \eta_{ik}\delta(\btheta - \btheta_i)$. 
	To push the algorithm to prioritize fitting edge effects for a certain number of iterations, we propose a variant of the M-step where we replace the right hand side with a \gls{KDE} that is unbiased at the boundary up to $\mathcal{O}(h)$, constructed with non-symmetric truncated Gaussian kernels $k(\btheta,\btheta') = \phi_{[\ba,\bb]}(\btheta \mid \bmu(\btheta'), {\bf h})$. 
	This \gls{KDE} is described in Appendix~\ref{sec:BoundaryKDE}, as is the function $\bmu(\cdot)$.
	With it, we set our estimate for $\hat{p}_i$ as
	\begin{equation}
		\hat{p}_i(\btheta) = \sum_{ik} \eta_{ik} \phi_{[\ba, \bb]}(\btheta \mid \bmu(\btheta_i), {\bf h})\,.
	\end{equation}
	Then our modified M-step is as follows.
	\textbf{M-step (\gls{KDE} variant): }For a certain number of iterations we perform a variant of the M-step update using 
	\begin{equation}
		\begin{aligned}
			\mathbb{E}_{\btheta \sim \mathcal{N}_{[\ba,\bb]}(\bmu_k, \bSigma_k)}[\btheta] &=  \mathbb{E}_{\btheta \sim \hat{p}_i(\btheta)}[\btheta] = \sum_i \eta_{ik} \mathcal{M}_{1 i} \\
			\mathbb{E}_{\btheta \sim \mathcal{N}_{[\ba,\bb]}(\bmu_k, \bSigma_k)}[\btheta\btheta^T] &=  \mathbb{E}_{\btheta \sim \hat{p}_i(\btheta)}[\btheta\btheta^T] = \sum_i \eta_{ik} \mathcal{M}_{2 i}\,,
		\end{aligned}
		\label{eq:moment-matching-updated-kde}
	\end{equation}	
	where $\mathcal{M}_{1i}$ and $\mathcal{M}_{2i}$ are the first and second moments of $ \mathcal{N}_{[\ba, \bb]}(\bmu(\btheta_i),\ {\bf h})$.
	We perform a certain number of EM iterations using this variant of the M-step and subsequently fall back to standard iterations as before. 
	Our full EM fitting procedure is summarized in Algorithm~\ref{alg:EM}.

	\section{Computational Techniques for the evalutation of the multivariate gaussian probability mass}
	
	\label{sec:ComputationOfIntegral}

	The function $C_{[\ba, \bb]}(\bmu, \bSigma)$ evaluates the probability mass of some multivariate gaussian $N(\bmu, \bSigma)$ within a hypercube truncation region defined by $[\ba, \bb]$. 
	This can be computed very quickly analytically for univariate gaussians and with a bit more compute for bivariate Gaussians \citep[see e.g.][]{Tsay2021, Figueiredo}. 
	With only these two implementations one can compute $C_{[\ba, \bb]}(\bmu, \bSigma)$ for covariance matrices of higher dimension that separate into blocks of at most $2\times 2$. 
	We have numerical routines to evaluate $C_{[\ba, \bb]}(\bmu, \bSigma)$ and $F$ 
	\added[id=AZ]{of Eq.~\eqref{eq:AnalyticIntegralF}} 
	in our package \href{https://www.github.com/gravpop}{\texttt{gravpop}} \citep{gravpop}. 
		
	With no structure on $\bSigma$ there is a psuedo-algorithm described in \cite{Genz1992} that can compute $C_{[\ba, \bb]}(\bmu, \bSigma)$ for general dimension. 
	This algorithm is implemented in the \texttt{Julia} package \texttt{MvNormalCDF.jl} \citep{MvNormalCDFjl}, and utilized in our package to fit Truncated Gaussian Mixtures \texttt{TruncatedGaussianMixtures.jl} 
	\citep{truncatedgaussianmixtures}. 
	
	The algorithm by \cite{Genz1992} for the case of a general covariance structure is not generally amenable to auto-differentiation routines, which we leverage in our population analysis code.
	This limits us to covariance structures consisting of $2\times 2$ blocks. 
	However, if future work requires it, one can write custom Jacobians by leveraging the fact that the first and second derivatives of $C_{[\ba, \bb]}(\bmu, \bSigma)$ can be related to the first and second moments of $\mathcal N_{[\ba, \bb]}(\bmu, \bSigma)$. 
	The computation of those moments is a crucial part of fitting \glspl{TGMM}, and we have implemented routines to compute them in \texttt{TruncatedGaussianMixtures.jl} \citep{truncatedgaussianmixtures}, based on the methods of \cite{Lee2012}.
	
	As with most estimates of CDFs, for very extreme features and when $\bmu$ moves far from the truncation region, numerical stability can become an issue. 
	We find that while the evaluation of $C_{[\ba, \bb]}(\bmu, \bSigma)$ can be stable in these situations, the gradients calculated by JAX can be unstable. 
	To get around this we perform a pre-processing step before population analysis, which maps Gaussian components in the \gls{TGMM} fit that are far from the truncation domain closer to the domain edge, without changing the distribution inside the domain very much. 
	The transformation
	\begin{equation}
		\begin{aligned}
			\alpha &= \frac{\mu - a}{\sigma}, \quad \beta = \frac{\mu - b}{\sigma}, \\
			\sigma' &= \begin{cases} 
				\sigma / \sqrt{n}, & \text{if component is} > 4\sigma \text{ outside domain} \\
				\sigma, & \text{otherwise},
			\end{cases} \\
			\mu' &= \begin{cases}
				a - \frac{a - \mu}{n}, & \text{if component is} > 4\sigma \text{ left of }a, \\
				b + \frac{\mu - b}{n}, & \text{if component is} > 4\sigma \text{ right of }b, \\
				\mu, & \text{otherwise}\,,
			\end{cases}
		\end{aligned}
	\end{equation}
	accomplishes this task, where $n$ controls the size of the effect. 
	We find that using $n=10$ made the derivatives stable and does not substantially affect inference.

	\section{Additional Results}
	\label{sec:AdditionalResults}
	
\begin{figure}[t]
		\begin{center}
			\includegraphics[width=0.8\paperwidth]{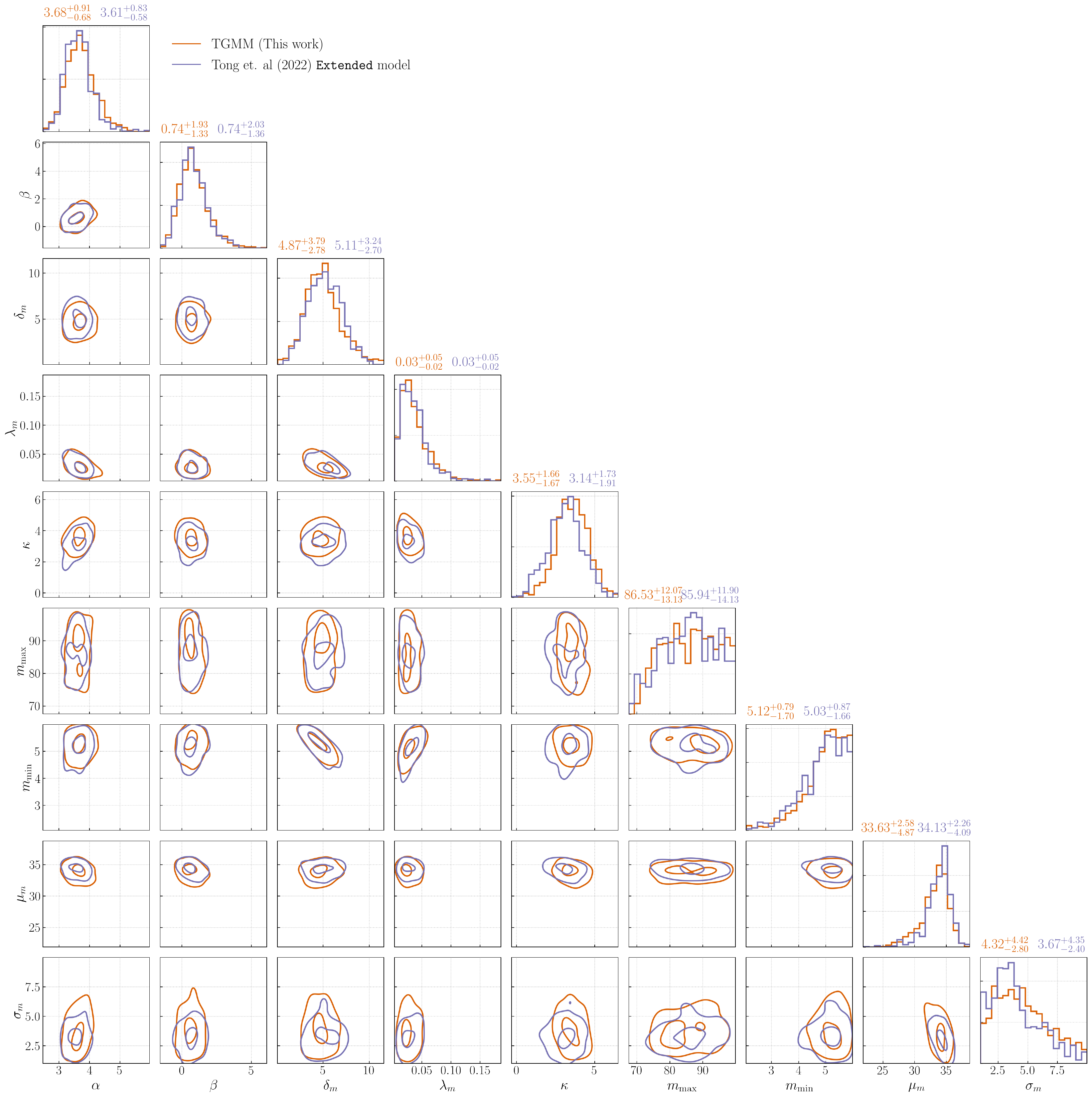}
		\end{center}
		\caption{As in Figure~\ref{fig:TongComparisonSpinMagnitude}, but depicting posterior distributions for the hyperparameters in our mass and redshift sectors.}
		\label{fig:TongComparisonRest}
	\end{figure}

	Here we provide further results for our full population inference.
	Figure~\ref{fig:TongComparisonRest} shows a corner plot comparing our \gls{TGMM}-based reproduction to the results of \cite{Tong2022}.
	We show the posteriors of the hyperparameters in the mass and redshift sector of the problem, finding good agreement across all parameters.

	\bibliographystyle{aasjournal}
	\bibliography{./pop_methods_paper.bbl}

\begin{thebibliography}{}
\expandafter\ifx\csname natexlab\endcsname\relax\def\natexlab#1{#1}\fi
\providecommand{\url}[1]{\href{#1}{#1}}
\providecommand{\dodoi}[1]{doi:~\href{http://doi.org/#1}{\nolinkurl{#1}}}
\providecommand{\doeprint}[1]{\href{http://ascl.net/#1}{\nolinkurl{http://ascl.net/#1}}}
\providecommand{\doarXiv}[1]{\href{https://arxiv.org/abs/#1}{\nolinkurl{https://arxiv.org/abs/#1}}}

\bibitem[{MvN(2021)}]{MvNormalCDFjl}
 2021, \bibinfo{title}{{G}it{H}ub - {P}harm{C}at/{M}v{N}ormal{C}{D}{F}.jl:
  {Q}uasi-{M}onte-{C}arlo numerical computation of multivariate normal
  probabilities --- github.com,},
  \url{https://github.com/PharmCat/MvNormalCDF.jl}

\bibitem[{J. Aasi {et~al.}(2015)Aasi {et~al.}}]{LIGOScientific:2014pky}
Aasi, J., {et~al.} 2015, \bibinfo{title}{{Advanced LIGO},} Class. Quant. Grav.,
  32, 074001, \dodoi{10.1088/0264-9381/32/7/074001}

\bibitem[{R. Abbott {et~al.}(2021)Abbott {et~al.}}]{LIGOScientific:2019lzm}
Abbott, R., {et~al.} 2021, \bibinfo{title}{{Open data from the first and second
  observing runs of Advanced LIGO and Advanced Virgo},} SoftwareX, 13, 100658,
  \dodoi{10.1016/j.softx.2021.100658}

\bibitem[{R. Abbott {et~al.}(2023{\natexlab{a}})Abbott
  {et~al.}}]{KAGRA:2023pio}
Abbott, R., {et~al.} 2023{\natexlab{a}}, \bibinfo{title}{{Open Data from the
  Third Observing Run of LIGO, Virgo, KAGRA, and GEO},} Astrophys. J. Suppl.,
  267, 29, \dodoi{10.3847/1538-4365/acdc9f}

\bibitem[{R. Abbott {et~al.}(2023{\natexlab{b}})Abbott
  {et~al.}}]{KAGRA:2021vkt}
Abbott, R., {et~al.} 2023{\natexlab{b}}, \bibinfo{title}{{GWTC-3: Compact
  Binary Coalescences Observed by LIGO and Virgo during the Second Part of the
  Third Observing Run},} Phys. Rev. X, 13, 041039,
  \dodoi{10.1103/PhysRevX.13.041039}

\bibitem[{R. Abbott {et~al.}(2023{\natexlab{c}})Abbott {et~al.}}]{LVKPop}
Abbott, R., {et~al.} 2023{\natexlab{c}}, \bibinfo{title}{Population of Merging
  Compact Binaries Inferred Using Gravitational Waves through GWTC-3,} Phys.
  Rev. X, 13, 011048, \dodoi{10.1103/PhysRevX.13.011048}

\bibitem[{R. Abbott {et~al.}(2024)Abbott {et~al.}}]{LIGOScientific:2021usb}
Abbott, R., {et~al.} 2024, \bibinfo{title}{{GWTC-2.1: Deep extended catalog of
  compact binary coalescences observed by LIGO and Virgo during the first half
  of the third observing run},} Phys. Rev. D, 109, 022001,
  \dodoi{10.1103/PhysRevD.109.022001}

\bibitem[{C. Adamcewicz {et~al.}(2025)Adamcewicz, Guttman, Lasky, \&
  Thrane}]{Adamcewicz2025-hm}
Adamcewicz, C., Guttman, N., Lasky, P.~D., \& Thrane, E. 2025,
  \bibinfo{title}{Do both black holes spin in merging binaries? Evidence from
  GWTC-4 and astrophysical implications,} arXiv,
  \dodoi{10.48550/ARXIV.2509.04706}

\bibitem[{S. Amari \& H. Nagaoka(2000)Amari \& Nagaoka}]{amari2000methods}
Amari, S., \& Nagaoka, H. 2000, Methods of Information Geometry (American
  Mathematical Society and Oxford University Press)

\bibitem[{A.~W. Bowman(1984)Bowman}]{crossvalidation-KDE}
Bowman, A.~W. 1984, \bibinfo{title}{An Alternative Method of Cross-Validation
  for the Smoothing of Density Estimates,} Biometrika, 71, 353.
\newblock \url{http://www.jstor.org/stable/2336252}

\bibitem[{T.~A. Callister {et~al.}(2024)Callister, Essick, \&
  Holz}]{Callister:2024qyq}
Callister, T.~A., Essick, R., \& Holz, D.~E. 2024, \bibinfo{title}{{Neural
  network emulator of the Advanced LIGO and Advanced Virgo selection
  function},} Phys. Rev. D, 110, 123041, \dodoi{10.1103/PhysRevD.110.123041}

\bibitem[{T.~A. Callister {et~al.}(2022)Callister, Miller, Chatziioannou, \&
  Farr}]{Callister:2022qwb}
Callister, T.~A., Miller, S.~J., Chatziioannou, K., \& Farr, W.~M. 2022,
  \bibinfo{title}{{No Evidence that the Majority of Black Holes in Binaries
  Have Zero Spin},} Astrophys. J. Lett., 937, L13,
  \dodoi{10.3847/2041-8213/ac847e}

\bibitem[{M.~F. Carney {et~al.}(2018)Carney, Wade, \& Irwin}]{Carney_2018}
Carney, M.~F., Wade, L.~E., \& Irwin, B.~S. 2018, \bibinfo{title}{Comparing two
  models for measuring the neutron star equation of state from
  gravitational-wave signals,} Physical Review D, 98,
  \dodoi{10.1103/physrevd.98.063004}

\bibitem[{R. Cloutier(2024)Cloutier}]{cloutier2024exoplanetl}
Cloutier, R. 2024, \bibinfo{title}{Exoplanet Demographics: Physical and Orbital
  Properties,} \doarXiv{2409.13062}

\bibitem[{T.~L.~S. Collaboration {et~al.}(2025{\natexlab{a}})Collaboration, the
  Virgo~Collaboration, \& the
  KAGRA~Collaboration}]{theligoscientificcollaboration2025gwtc40populationpropertiesmerging}
Collaboration, T. L.~S., the Virgo~Collaboration, \& the KAGRA~Collaboration.
  2025{\natexlab{a}}, \bibinfo{title}{GWTC-4.0: Population Properties of
  Merging Compact Binaries,} \doarXiv{2508.18083}

\bibitem[{T.~L.~S. Collaboration {et~al.}(2025{\natexlab{b}})Collaboration, the
  Virgo~Collaboration, \& the
  KAGRA~Collaboration}]{theligoscientificcollaboration2025gwtc40updatinggravitationalwavetransient}
Collaboration, T. L.~S., the Virgo~Collaboration, \& the KAGRA~Collaboration.
  2025{\natexlab{b}}, \bibinfo{title}{GWTC-4.0: Updating the Gravitational-Wave
  Transient Catalog with Observations from the First Part of the Fourth
  LIGO-Virgo-KAGRA Observing Run,} \doarXiv{2508.18082}

\bibitem[{T.~L.~S. Collaboration {et~al.}(2025{\natexlab{c}})Collaboration, the
  Virgo~Collaboration, \& the
  KAGRA~Collaboration}]{theligoscientificcollaboration2025gwtc40methodsidentifyingcharacterizing}
Collaboration, T. L.~S., the Virgo~Collaboration, \& the KAGRA~Collaboration.
  2025{\natexlab{c}}, \bibinfo{title}{GWTC-4.0: Methods for Identifying and
  Characterizing Gravitational-wave Transients,} \doarXiv{2508.18081}

\bibitem[{V. Delfavero {et~al.}(2022{\natexlab{a}})Delfavero, O'Shaughnessy,
  Wysocki, \&
  Yelikar}]{delfavero2022compressedparametricnonparametricapproximations}
Delfavero, V., O'Shaughnessy, R., Wysocki, D., \& Yelikar, A.
  2022{\natexlab{a}}, \bibinfo{title}{Compressed Parametric and Non-Parametric
  Approximations to the Gravitational Wave Likelihood,} \doarXiv{2205.14154}

\bibitem[{V. Delfavero {et~al.}(2022{\natexlab{b}})Delfavero, O'Shaughnessy,
  Wysocki, \& Yelikar}]{delfavero2022normalapproximatelikelihoodsgravitational}
Delfavero, V., O'Shaughnessy, R., Wysocki, D., \& Yelikar, A.
  2022{\natexlab{b}}, \bibinfo{title}{Normal Approximate Likelihoods to
  Gravitational Wave Events,} \doarXiv{2107.13082}

\bibitem[{B. Edelman {et~al.}(2023)Edelman, Farr, \& Doctor}]{Edelman_2023}
Edelman, B., Farr, B., \& Doctor, Z. 2023, \bibinfo{title}{Cover Your Basis:
  Comprehensive Data-driven Characterization of the Binary Black Hole
  Population,} The Astrophysical Journal, 946, 16,
  \dodoi{10.3847/1538-4357/acb5ed}

\bibitem[{B. {Edelman} {et~al.}(2023){Edelman}, {Farr}, \&
  {Doctor}}]{git_gwinferno}
{Edelman}, B., {Farr}, B., \& {Doctor}, Z. 2023,
  \bibinfo{title}{Gravitational-Wave Hierarchical Inference with NumPyro,}
  \url{https://github.com/FarrOutLab/GWInferno}

\bibitem[{R. Essick(2023)Essick}]{essick_semianalytic_2023}
Essick, R. 2023, \bibinfo{title}{Semianalytic sensitivity estimates for
  catalogs of gravitational-wave transients,} Physical Review D, 108, 043011,
  \dodoi{10.1103/PhysRevD.108.043011}

\bibitem[{R. Essick \& M. Fishbach(2024)Essick \& Fishbach}]{Essick:2023upv}
Essick, R., \& Fishbach, M. 2024, \bibinfo{title}{{Ensuring Consistency between
  Noise and Detection in Hierarchical Bayesian Inference},} Astrophys. J., 962,
  169, \dodoi{10.3847/1538-4357/ad1604}

\bibitem[{W.~M. Farr(2019)Farr}]{Farr:2019rap}
Farr, W.~M. 2019, \bibinfo{title}{{Accuracy Requirements for
  Empirically-Measured Selection Functions},} Research Notes of the AAS, 3, 66,
  \dodoi{10.3847/2515-5172/ab1d5f}

\bibitem[{F. Figueiredo(2023)Figueiredo}]{Figueiredo}
Figueiredo, F. 2023, \bibinfo{title}{JAX Implementation of binormalCDF,} Flavio
  Figueiredo.
\newblock
  \url{https://colab.research.google.com/drive/1w2tI1-1LWzPSdG_jE0FXzdrs6VAsJwhv?usp=sharing}

\bibitem[{D. {Foreman-Mackey} {et~al.}(2013){Foreman-Mackey}, Hogg, Lang, \&
  Goodman}]{foreman-mackeyEmceeMCMCHammer2013}
{Foreman-Mackey}, D., Hogg, D.~W., Lang, D., \& Goodman, J. 2013,
  \bibinfo{title}{Emcee : {{The MCMC Hammer}},} Publications of the
  Astronomical Society of the Pacific, 125, 306, \dodoi{10.1086/670067}

\bibitem[{D. Frisch \& U.~D. Hanebeck(2021)Frisch \&
  Hanebeck}]{frisch2021gaussianmixtureestimationweighted}
Frisch, D., \& Hanebeck, U.~D. 2021, \bibinfo{title}{Gaussian Mixture
  Estimation from Weighted Samples,} \doarXiv{2106.05109}

\bibitem[{S. Galaudage {et~al.}(2021)Galaudage, Talbot, Nagar, Jain, Thrane, \&
  Mandel}]{Galaudage2021}
Galaudage, S., Talbot, C., Nagar, T., {et~al.} 2021, \bibinfo{title}{Building
  Better Spin Models for Merging Binary Black Holes: Evidence for Nonspinning
  and Rapidly Spinning Nearly Aligned Subpopulations,} The Astrophysical
  Journal Letters, 921, L15, \dodoi{10.3847/2041-8213/ac2f3c}

\bibitem[{A. Genz(1992)Genz}]{Genz1992}
Genz, A. 1992, \bibinfo{title}{Numerical Computation of Multivariate Normal
  Probabilities,} Journal of Computational and Graphical Statistics, 1,
  141–149, \dodoi{10.1080/10618600.1992.10477010}

\bibitem[{J. Heinzel {et~al.}(2025)Heinzel, Mould, Álvarez López, \&
  Vitale}]{Heinzel_2025}
Heinzel, J., Mould, M., Álvarez López, S., \& Vitale, S. 2025,
  \bibinfo{title}{High resolution nonparametric inference of gravitational-wave
  populations in multiple dimensions,} Physical Review D, 111,
  \dodoi{10.1103/physrevd.111.063043}

\bibitem[{D.~W. Hogg {et~al.}(2010)Hogg, Myers, \& Bovy}]{hogg_inferring_2010}
Hogg, D.~W., Myers, A.~D., \& Bovy, J. 2010, \bibinfo{title}{{INFERRING} {THE}
  {ECCENTRICITY} {DISTRIBUTION},} The Astrophysical Journal, 725, 2166,
  \dodoi{10.1088/0004-637X/725/2/2166}

\bibitem[{A. Hussain(2024{\natexlab{a}})Hussain}]{truncatedgaussianmixtures}
Hussain, A. 2024{\natexlab{a}},
  \bibinfo{title}{\textbf{truncatedgaussianmixtures}: Fitting truncated
  Gaussian mixture models,} Zenodo, \dodoi{10.5281/zenodo.13999561}

\bibitem[{A. Hussain(2024{\natexlab{b}})Hussain}]{gravpop}
Hussain, A. 2024{\natexlab{b}}, \bibinfo{title}{\textbf{gravpop}: Astrophysical
  population modeling for gravitational waves with the ability to probe narrow
  population features over bounded domains,} Zenodo,
  \dodoi{10.5281/zenodo.14003052}

\bibitem[{A. Hussain {et~al.}(2024)Hussain, Isi, \&
  Zimmerman}]{hussainHintsSpinmagnitudeCorrelations2024a}
Hussain, A., Isi, M., \& Zimmerman, A. 2024, \bibinfo{title}{Hints of
  Spin-Magnitude Correlations and a Rapidly Spinning Subpopulation of Binary
  Black Holes,} arXiv, \dodoi{10.48550/ARXIV.2411.02252}

\bibitem[{D. Keating \& N.~B. Cowan(2021)Keating \&
  Cowan}]{keating_atmospheric_2021}
Keating, D., \& Cowan, N.~B. 2021, \bibinfo{title}{Atmospheric characterization
  of hot Jupiters using hierarchical models of \textit{Spitzer} observations,}
  Monthly Notices of the Royal Astronomical Society, 509, 289,
  \dodoi{10.1093/mnras/stab2941}

\bibitem[{B.~D. Lackey \& L. Wade(2015)Lackey \& Wade}]{Lackey_2015}
Lackey, B.~D., \& Wade, L. 2015, \bibinfo{title}{Reconstructing the
  neutron-star equation of state with gravitational-wave detectors from a
  realistic population of inspiralling binary neutron stars,} Physical Review
  D, 91, \dodoi{10.1103/physrevd.91.043002}

\bibitem[{G. Lee \& C. Scott(2012)Lee \& Scott}]{Lee2012}
Lee, G., \& Scott, C. 2012, \bibinfo{title}{EM algorithms for multivariate
  Gaussian mixture models with truncated and censored data,} Computational
  Statistics \& Data Analysis, 56, 2816–2829,
  \dodoi{10.1016/j.csda.2012.03.003}

\bibitem[{B. Leistedt {et~al.}(2023)Leistedt, Alsing, Peiris, Mortlock, \&
  Leja}]{Leistedt_2023}
Leistedt, B., Alsing, J., Peiris, H., Mortlock, D., \& Leja, J. 2023,
  \bibinfo{title}{Hierarchical Bayesian Inference of Photometric Redshifts with
  Stellar Population Synthesis Models,} The Astrophysical Journal Supplement
  Series, 264, 23, \dodoi{10.3847/1538-4365/ac9d99}

\bibitem[{ {LIGO Scientific, Virgo and KAGRA Collaborations}(2021){LIGO
  Scientific, Virgo and KAGRA Collaborations}}]{LIGO-SearchSensitivity}
{LIGO Scientific, Virgo and KAGRA Collaborations}. 2021,
  \bibinfo{title}{{GWTC-3: Compact Binary Coalescences Observed by LIGO and
  Virgo During the Second Part of the Third Observing Run — O1+O2+O3 Search
  Sensitivity Estimates},} Zenodo, \dodoi{10.5281/zenodo.5636816}

\bibitem[{ {{LIGO Scientific}, {Virgo,} and {KAGRA} Collaborations}(2024){{LIGO
  Scientific}, {Virgo,} and {KAGRA} Collaborations}}]{GWOSC}
{{LIGO Scientific}, {Virgo,} and {KAGRA} Collaborations}. 2024,
  \bibinfo{title}{{Gravitational Wave Open Science Center},},
  \url{https://www.gw-openscience.org/}

\bibitem[{S. Lloyd(1982)Lloyd}]{Lloyd:1982zni}
Lloyd, S. 1982, \bibinfo{title}{{Least squares quantization in PCM},} IEEE
  Trans. Info. Theor., 28, 129, \dodoi{10.1109/TIT.1982.1056489}

\bibitem[{J. Lustig-Yaeger {et~al.}(2022)Lustig-Yaeger, Sotzen, Stevenson,
  Luger, May, Mayorga, Mandt, \& Izenberg}]{lustig-yaeger_hierarchical_2022}
Lustig-Yaeger, J., Sotzen, K.~S., Stevenson, K.~B., {et~al.} 2022,
  \bibinfo{title}{Hierarchical Bayesian Atmospheric Retrieval Modeling for
  Population Studies of Exoplanet Atmospheres: A Case Study on the Habitable
  Zone,} The Astronomical Journal, 163, 140, \dodoi{10.3847/1538-3881/ac5034}

\bibitem[{M. Mancarella \& D. Gerosa(2025)Mancarella \&
  Gerosa}]{mancarella2025samplinghierarchicalpopulationposterior}
Mancarella, M., \& Gerosa, D. 2025, \bibinfo{title}{Sampling the full
  hierarchical population posterior distribution in gravitational-wave
  astronomy,} \doarXiv{2502.12156}

\bibitem[{I. Mandel {et~al.}(2019)Mandel, Farr, \& Gair}]{Mandel:2018mve}
Mandel, I., Farr, W.~M., \& Gair, J.~R. 2019, \bibinfo{title}{{Extracting
  distribution parameters from multiple uncertain observations with selection
  biases},} Mon. Not. Roy. Astron. Soc., 486, 1086,
  \dodoi{10.1093/mnras/stz896}

\bibitem[{J.~S. Marron \& D. Ruppert(1994)Marron \&
  Ruppert}]{marronTransformationsReduceBoundary1994}
Marron, J.~S., \& Ruppert, D. 1994, \bibinfo{title}{Transformations to {{Reduce
  Boundary Bias}} in {{Kernel Density Estimation}},} Journal of the Royal
  Statistical Society Series B: Statistical Methodology, 56, 653,
  \dodoi{10.1111/j.2517-6161.1994.tb02006.x}

\bibitem[{S. {Mastrogiovanni} {et~al.}(2023){Mastrogiovanni}, {Pierra},
  {Perri{\`e}s}, {Laghi}, {Santoro}, {Ghosh}, {Gray}, {Karathanasis}, \&
  {Leyde}}]{git_icarogw}
{Mastrogiovanni}, S., {Pierra}, G., {Perri{\`e}s}, S., {et~al.} 2023,
  \bibinfo{title}{ICAROGW Pure python package to estimate population properties
  of noisy observations,}
  \url{https://github.com/simone-mastrogiovanni/icarogw}

\bibitem[{S. {Mastrogiovanni} {et~al.}(2024){Mastrogiovanni}, {Pierra},
  {Perri{\`e}s}, {Laghi}, {Santoro}, {Ghosh}, {Gray}, {Karathanasis}, \&
  {Leyde}}]{icarogw_paper}
{Mastrogiovanni}, S., {Pierra}, G., {Perri{\`e}s}, S., {et~al.} 2024,
  \bibinfo{title}{{ICAROGW: A python package for inference of astrophysical
  population properties of noisy, heterogeneous, and incomplete observations},}
  \aap, 682, A167, \dodoi{10.1051/0004-6361/202347007}

\bibitem[{R.~O. Meesum~Qazalbash(2024)Meesum~Qazalbash}]{git_gwkokab}
Meesum~Qazalbash, Muhammad~Zeeshan, R.~O. 2024, \bibinfo{title}{{GWKokab}: A
  JAX-based gravitational-wave population inference toolkit,}, 0.0.1
  \url{https://github.com/gwkokab/gwkokab}

\bibitem[{N. Metropolis {et~al.}(1953)Metropolis, Rosenbluth, Rosenbluth,
  Teller, \& Teller}]{metropolisEquationStateCalculations1953}
Metropolis, N., Rosenbluth, A.~W., Rosenbluth, M.~N., Teller, A.~H., \& Teller,
  E. 1953, \bibinfo{title}{Equation of {{State Calculations}} by {{Fast
  Computing Machines}},} The Journal of Chemical Physics, 21, 1087,
  \dodoi{10.1063/1.1699114}

\bibitem[{G. Pratten {et~al.}(2021)Pratten {et~al.}}]{Pratten:2020ceb}
Pratten, G., {et~al.} 2021, \bibinfo{title}{{Computationally efficient models
  for the dominant and subdominant harmonic modes of precessing binary black
  holes},} Phys. Rev. D, 103, 104056, \dodoi{10.1103/PhysRevD.103.104056}

\bibitem[{M. Qazalbash {et~al.}(2025)Qazalbash, Zeeshan, \&
  O'Shaughnessy}]{qazalbash2025gwkokabimplementationidentifyproperties}
Qazalbash, M., Zeeshan, M., \& O'Shaughnessy, R. 2025, \bibinfo{title}{GWKokab:
  An Implementation to Identify the Properties of Multiple Population of
  Gravitational Wave Sources,} \doarXiv{2509.13638}

\bibitem[{C.~P. Robert \& G. Casella(2004)Robert \&
  Casella}]{robert_monte_2004}
Robert, C.~P., \& Casella, G. 2004, Monte {Carlo} {Statistical} {Methods},
  Springer {Texts} in {Statistics} (New York, NY: Springer New York),
  \dodoi{10.1007/978-1-4757-4145-2}

\bibitem[{L.~A. Rogers(2015)Rogers}]{rogers_most_2015}
Rogers, L.~A. 2015, \bibinfo{title}{\textit{{MOST}} 1.6 {EARTH}-{RADIUS}
  {PLANETS} {ARE} {NOT} {ROCKY},} The Astrophysical Journal, 801, 41,
  \dodoi{10.1088/0004-637X/801/1/41}

\bibitem[{E.~F.
  Schuster(1985)Schuster}]{schusterIncorporatingSupportConstraints1985}
Schuster, E.~F. 1985, \bibinfo{title}{Incorporating support constraints into
  nonparametric estimators of densities,} Communications in Statistics - Theory
  and Methods, 14, 1123, \dodoi{10.1080/03610928508828965}

\bibitem[{B.
  Silverman(2018)Silverman}]{silvermanDensityEstimationStatistics2018}
Silverman, B. 2018, Density {Estimation} for {Statistics} and {Data}
  {Analysis}, 1st edn. (Routledge), \dodoi{10.1201/9781315140919}

\bibitem[{J. Skilling(2006)Skilling}]{skillingNestedSamplingGeneral2006}
Skilling, J. 2006, \bibinfo{title}{Nested Sampling for General {{Bayesian}}
  Computation,} Bayesian Analysis, 1, \dodoi{10.1214/06-BA127}

\bibitem[{C. Talbot {et~al.}(2025)Talbot, Farah, Galaudage, Golomb, \&
  Tong}]{Talbot2025}
Talbot, C., Farah, A., Galaudage, S., Golomb, J., \& Tong, H. 2025,
  \bibinfo{title}{GWPopulation: Hardware agnostic population inference for
  compact binaries and beyond,} Journal of Open Source Software, 10, 7753,
  \dodoi{10.21105/joss.07753}

\bibitem[{C. Talbot \& J. Golomb(2023)Talbot \& Golomb}]{Talbot2023}
Talbot, C., \& Golomb, J. 2023, \bibinfo{title}{Growing pains: understanding
  the impact of likelihood uncertainty on hierarchical Bayesian inference for
  gravitational-wave astronomy,} Monthly Notices of the Royal Astronomical
  Society, 526, 3495–3503, \dodoi{10.1093/mnras/stad2968}

\bibitem[{C. Talbot \& E. Thrane(2022)Talbot \& Thrane}]{Talbot:2020oeu}
Talbot, C., \& Thrane, E. 2022, \bibinfo{title}{{Flexible and Accurate
  Evaluation of Gravitational-wave Malmquist Bias with Machine Learning},}
  Astrophys. J., 927, 76, \dodoi{10.3847/1538-4357/ac4bc0}

\bibitem[{E. Thrane \& C. Talbot(2019)Thrane \& Talbot}]{Thrane:2018qnx}
Thrane, E., \& Talbot, C. 2019, \bibinfo{title}{{An introduction to Bayesian
  inference in gravitational-wave astronomy: parameter estimation, model
  selection, and hierarchical models},} Publ. Astron. Soc. Austral., 36, e010,
  \dodoi{10.1017/pasa.2019.2}

\bibitem[{H. Tong {et~al.}(2022)Tong, Galaudage, \& Thrane}]{Tong2022}
Tong, H., Galaudage, S., \& Thrane, E. 2022, \bibinfo{title}{Population
  properties of spinning black holes using the gravitational-wave transient
  catalog 3,} Physical Review D, 106, \dodoi{10.1103/physrevd.106.103019}

\bibitem[{W.-J. Tsay \& P.-H. Ke(2021)Tsay \& Ke}]{Tsay2021}
Tsay, W.-J., \& Ke, P.-H. 2021, \bibinfo{title}{A simple approximation for the
  bivariate normal integral,} Communications in Statistics - Simulation and
  Computation, 52, 1462–1475, \dodoi{10.1080/03610918.2021.1884718}

\bibitem[{P. Virtanen {et~al.}(2020)Virtanen, Gommers, Oliphant, Haberland,
  Reddy, Cournapeau, Burovski, Peterson, Weckesser, Bright, {van der Walt},
  Brett, Wilson, Millman, Mayorov, Nelson, Jones, Kern, Larson, Carey, Polat,
  Feng, Moore, {VanderPlas}, Laxalde, Perktold, Cimrman, Henriksen, Quintero,
  Harris, Archibald, Ribeiro, Pedregosa, {van Mulbregt}, \& {SciPy 1.0
  Contributors}}]{2020SciPy-NMeth}
Virtanen, P., Gommers, R., Oliphant, T.~E., {et~al.} 2020,
  \bibinfo{title}{{{SciPy} 1.0: Fundamental Algorithms for Scientific Computing
  in Python},} Nature Methods, 17, 261, \dodoi{10.1038/s41592-019-0686-2}

\bibitem[{A. Wolfgang {et~al.}(2016)Wolfgang, Rogers, \&
  Ford}]{wolfgang_probabilistic_2016}
Wolfgang, A., Rogers, L.~A., \& Ford, E.~B. 2016,
  \bibinfo{title}{{PROBABILISTIC} {MASS}–{RADIUS} {RELATIONSHIP} {FOR}
  {SUB}-{NEPTUNE}-{SIZED} {PLANETS},} The Astrophysical Journal, 825, 19,
  \dodoi{10.3847/0004-637X/825/1/19}

\bibitem[{D. Wysocki {et~al.}(2020)Wysocki, O'Shaughnessy, Wade, \&
  Lange}]{wysocki2020inferringneutronstarequation}
Wysocki, D., O'Shaughnessy, R., Wade, L., \& Lange, J. 2020,
  \bibinfo{title}{Inferring the neutron star equation of state simultaneously
  with the population of merging neutron stars,} \doarXiv{2001.01747}

\end{thebibliography}
	
\end{document}